\documentclass[debug]{rmaa}

\usepackage{graphicx}	% Including figure files
\usepackage{amsmath}	% Advanced maths commands
\usepackage{multirow}
\usepackage{tabularx}
\usepackage{spverbatim}
\usepackage{makecell}
\usepackage{hyperref}
\usepackage{longtable}

\usepackage{paralist}

\usepackage{psfrag,color}

\usepackage[latin1]{inputenc}

\newcommand{\xs}{$\mathtt{XookSuut}$}
\newcommand{\diskfit}{$\mathtt{DiskFit}$}
\newcommand{\kms}{km\,s$^{-1}$}
\newcommand{\PAdisk}{$\phi_\mathrm{disk}^\prime$}
\newcommand{\PAbar}{$\phi_\mathrm{bar}$}

\newcommand{\SN}{S$/$N}
\newcommand{\hi}{\ion{H}{i}}
\newcommand{\hii}{\ion{H}{ii}}

\newcommand{\ha}{H$\alpha$}

\title{\xs: a bayesian tool for modeling circular and non--circular flows on 2D velocity maps}

\author{
  C. L\'opez-Cob\'a,\altaffilmark{1} 
  Lihwai Lin,\altaffilmark{1}
  Sebasti\'an~F.~S\'anchez\altaffilmark{2}}

\altaffiltext{1}{Institute of Astronomy \& Astrophysics, Academia Sinica, 106, Taipei, Taiwan.}

\altaffiltext{2}{Instituto de Astronom\'ia, Universidad Nacional Aut\'onoma de  M\'exico \\
Circuito Exterior, Ciudad Universitaria, Ciudad de M\'exico 04510,  Mexico.}

\shortauthor{L\'opez-Cob\'a et al.}
\shorttitle{\xs}

\fulladdresses{
\item Carlos~L\'opez-Cob\'a: Institute of Astronomy \& Astrophysics, Academia Sinica, 106, Taipei, Taiwan (calopez@asiaa.sinica.edu.tw).
\item Lihwai Lin : Institute of Astronomy \& Astrophysics, Academia Sinica, 106, Taipei, Taiwan (lihwailin@asiaa.sinica.edu.tw)
\item Sebasti\'an~F.~S\'anchez : Instituto de Astronom\'ia, Universidad Nacional Aut\'onoma de  M\'exico.
Circuito Exterior, Ciudad Universitaria, Ciudad de M\'exico 04510,  Mexico (sfsanchez@astro.unam.mx)
}

\listofauthors{C.~L\'opez-Cob\'a , Lihwai Lin, \& Sebasti\'an~F.~S\'anchez}
\indexauthor{L\'opez-Cob\'a, C.}
\indexauthor{Lihwai, L.}
\indexauthor{S\'anchez, S. F.}

\abstract{ We present \xs, a Python implementation of the \texttt{DiskFit} algorithm, optimized to perform robust Bayesian inference on
parameters describing models of circular and noncircular rotation in galaxies. \xs~surges as a Bayesian alternative for kinematic modeling of 2D velocity maps;
it implements efficient sampling methods, specifically Markov Chain Monte Carlo (MCMC) and Nested Sampling (NS), to obtain the posteriors and marginalized distributions of kinematic models including circular motions, axisymmetric radial flows, bisymmetric flows, and harmonic decomposition of the LoS~velocity.
In this way, kinematic models are obtained by pure sampling methods, rather than standard minimization techniques based on the $\chi^2$.
All together, \xs~ represents a sophisticated tool for deriving rotational curves and to explore the error distribution and covariance between parameters.
}

\resumen{
En este art\'iculo presentamos \xs, una implenetaci\'on en Python del algoritmo \diskfit,
optimizado para realizar inferencia Bayesiana de manera robusta sobre par\'ametros que describen
modelos cinem\'aticos de rotaci\'on circular y no circular an galaxias. \xs~surge como una
alternativa Bayesiana para el modelado cinem\'atico de mapas dos dimensionales; el c\'odigo 
implementa m\'etodos de muestreo eficiente, espec\'ificamente Markov Chain Monte Carlo y Nested Sampling, para
obtener las distribuciones posteriores y marginalizadas de modelos cinem\'aticos entre los que se encuentran:
movimientos circulares, flujos radiales axisim\'etricos, flujos bisim\'etricos y una descomposici\'on en harm\'onicos
del campo de velocidad. De esta manera, los modelos cinem\'aticos son derivados por m\'etodos de muestreo en vez
de adoptar t\'ecnicas de minimizaci\'on basados en la $\chi^2$. Finalmente, \xs~representa una heramienta sofisticada para derivar
curvas de rotaci\'on y explorar la distribuci\'on de errores y covarianza entre par\'ametros. }

\addkeyword{galaxies: kinematics and dynamics}
\addkeyword{software: data analysis}

\begin{document}
% Typeset article header
\maketitle

%%%%%%%%%%%%%%%%%%%%%%%%%%%%%%%%%%%%%%%%%%%%%%%%%%

%%%%%%%%%%%%%%%%% BODY OF PAPER %%%%%%%%%%%%%%%%%%

\section{Introduction} \label{sec:intro}

The rotation pattern observed on two-dimensional velocity maps is the result of the gravitational potential and the mass distribution in a galaxy  together with environmental factors and projection effects \citep[e.g.,][]{Rubin1970,Binney}. In disk--like systems the rotation, or azimuthal velocity, is the dominant velocity component. When this velocity is plotted against the galactocentric distance it describes the rotational curve of a galaxy \citep[e.g.][]{Rubin1970,Rubin1980}. 

Since early studies of the neutral hydrogen distribution on nearby galaxies, it was possible to obtain resolved velocity fields \citep[e.g.,][]{Warner1973}; these \hi~velocity maps showed ordered kinematic patterns that in most cases could be described by pure circular rotation \citep[e.g.,][]{Wright1971,Begeman1989,THINGS}.
Since then, many efforts have been done for recovering rotation curves of galaxies, not only in \hi~data, but also in molecular and ionized gas observations. \citet{Begeman1987PhD,Begeman1989} introduced 
a methodology to extract the rotational velocity curve from two-dimensional (2D) velocity maps based on the so--called tilted rings.
This idea became the core of most of the algorithms focused on the determination of the rotation curves of galaxies, for instance the \texttt{GIPSY} task \texttt{ROTCUR}~\citep[e.g.,][]{Begeman1987PhD}.
The tilted ring model assumes that the observed velocity field can be described by pure circular motions with possible variations in the projection angles.
From it, several algorithms have been developed to study the kinematic structures of galaxies. On one side are those who uses 3D data-cubes, for instance   {\tt 3DBarolo}~\citep[e.g.,][]{3Dbarolo},
{\tt TiRiFiC } \citep[e.g.,][]{TiRiFiC}, {\tt GalPak$^{3D}$} \citep[e.g.,][]{Galpack}, {\tt KinMSpy} \citep[][]{kinms}.
In a second category are those which work on velocity fields, such as {\tt RESWRI} \citep[e.g.,][]{RESWRI}, \texttt{DiskFit} \citep[e.g.,][]{diskfit}, {\tt 2DBAT} \citep[e.g.,][]{2DBAT}, and \texttt{KINEMETRY} \citep[e.g.,][]{Kinemetry} among others.

3D algorithms have the advantages of extracting all the information from the datacubes. These methods model the entire datacubes, which allows them to correct for beam smearing effects and also to handle projection effects. However, the inclusion of datacubes usually involves the addition of extra parameters during the fitting process, which in most cases, involves larger computing time depending on the dimensions of the datacube and the fitting routine. On the other hand, 2D algorithms work on the projected line of sight velocity (LOSV); for this reason they tend to be faster than 3D methods. If galaxies are not severely affected by spatial resolution effects, (i.e., the observational point spread function, PSF), both methods show consistent results in rotational velocities \citep[e.g.,][]{Kamphuis2015}. 

Nevertheless, non--circular motions driven by structural components of galaxies (such as spiral arms, bars, bulge), or by angular momentum lost,  are not included within the circular rotation assumption \citep[e.g.,][]{Kormendy1983, Lacey1985, Wong2004}; 
nor those motions induced by internal processes (stellar winds, \hii~regions, shocks, outflows). Altogether, and taking into account  projection effects, make the modeling of non--circular motions a big challenge.
Only a few algorithms take into account deviation of circular motions, among which are: {\tt TiRiFiC}, ideal for modeling warp disks; \texttt{DiskFit}, suitable to model bar-like and radial flows; and \texttt{KINEMETRY} models non--circular motions of any order through harmonic decomposition. 

For deriving rotational curves, most algorithms adopt frequentist methods that minimize the residuals from a model function and the data, and those parameters that minimize the residuals are chosen for creating the kinematic model that better describes the data.
This means that from a frequentist perspective, there is
a single set of true parameters that describe the data.
Conversely, Bayesian methods assume that model parameters are totally random variables, and each parameter has associated a probability density function. In this way solutions are based on the likelihood of a parameter given the data; that is, on the posterior distribution of the parameters.

These are two different perspectives to estimate the parameters from a model.
Rotational curves are often described by several parameters, which makes it a high-dimensional problem, and therefore susceptible to find local solutions. Therefore, it is worth exploring methods that survey the parameter space of kinematic models to derive the best representation of the observed rotation patterns of disk galaxies.

In this paper we introduce \xs\footnote{\url{https://github.com/CarlosCoba/XookSuut-code}}(or \texttt{XS} for short). This is a Python tool that implements Bayesian methods for modeling circular and noncircular motions on 2D velocity maps.  
%
%\texttt{DiskFit} models and Fourier decomposition for extracting non--circular rotation patterns on velocity fields.
%
%\xs~adopts the \texttt{DiskFit} philosophy with a Bayesian approach and expand the non-circular models to any harmonic decomposition order. The Python ambient makes it easy to distribute and install. Moreover it is easy to systematize for its use in large galaxy samples, such as integral field spectroscopy data which offer kinematic properties of a wide variety of emission lines.
%
The name of this tool is a combination of two Mayan words:  \texttt{Xook} which means ``study'' and \texttt{Suut} which means ``rotation''.

This paper is organized as follows. In Section 2 we describe the different kinematic models included in \xs. In Section 3 we describe the algorithm,  the fitting procedure, and the error estimations. In Section 4 we show the performance of this code when it is applied on simulated velocity fields of galaxies with oval distortions, as well as on real velocity maps. Finally, in Section 6 we present our conclusions.

\section{Kinematic models}
\label{sec:kinematic_models}
In this section we describe the kinematic models included in \xs. We start with the simplest model, which is the circular rotation model, then we add a radial term for modeling radial flows. A bisymmetric model is included for describing oval distortions (i.e., bar-driven flows); finally, \xs~includes a more general harmonic decomposition of the line of sight velocity, for a total of  three non--circular rotation models.
For constructing these models, \xs~assumes that galaxies are flat and circular systems and they are viewed in projection, with a constant position angle ($\phi_\mathrm{disk}^\prime$)\footnote{Angles measured in the sky plane are marked with a prime symbol $(^\prime)$, otherwise they are measured in the galaxy plane. The disk position angle is measured from the north to east for the receding side of the galaxy.}, fixed inclination ($i$), fixed kinematic center ($x_0, y_0$) and constant systemic velocity throughout the disk. 
The flat disk approximation represents the more suitable assumption whenever the spatial resolution of the data, i.e., the point spread function, dominates over the typical thickness of disks.
With these assumptions, galaxies with strong warped disks are excluded. In addition,  systems where the inclination or position angle varies as a function of the galactocentric distance are also excluded since radial variations in these angles induce artificial non-circular motions when observed in projection, and such contribution to the line-of-sight velocities would be difficult to discern from  true non-circular motions \citep[e.g.,][]{Schoenmakers1997}.

\xs~adopts the methodology introduced by {\tt DiskFit} \citep[e.g.,][]{diskfit}~for creating a two dimensional interpolated map of the referred kinematic models and described in the following sections.

\subsection{Circular model}
\label{sec:circular_model}
The simplest model included in \xs~is the circular rotation model, which is the most frequently adopted for describing the rotation of galaxies. It assumes no other movements than pure circular motions in the plane of the disk and describes the rotation curve of disk galaxies.

Assuming that particles follow circular orbits on the disk, the circular model is given  by the projection of the velocity vector $\vec{V}$ along the line--of--sight direction:
\begin{equation}
 \label{circular}
 V_\mathrm{circ,model} = V_\mathrm{sys} + V_t(r)\sin i\cos \theta 
\end{equation}
$V_t$ is the circular rotation or azimuthal velocity and is a function of the galactocentric distance; $V_\mathrm{sys}$ is the systemic velocity and is assumed constant for all points in the galaxy. In this equation and in the following, $r$ is the radius of a circle in the disk plane, which projects to an ellipse in the sky plane. The angle $\theta$~is the azimuthal angle relative to the disk major axis, and $i$ is the disk inclination angle.

\subsection{Radial model}
When radial motions are not negligible, the disk circular velocity is described by two components of the velocity vector: the tangential velocity $(V_t)$ and the radial one $(V_r)$.
In this way, the model including radial velocities is described by the following expression:
\begin{equation}
 \label{radial}
 V_\mathrm{rad,model} = V_\mathrm{sys} + \sin i (V_t(r)\cos \theta  + V_r(r) \sin \theta ) 
\end{equation}
Comparing with Eq.~\ref{circular}, the only difference is the addition of the  $V_r \sin \theta$ term. This term accounts for axisymmetric radial flows (inflow or outflow) on the disk plane.

\subsection{Bissymetric model}
The bisymmetric model describes an oval distortion on the velocity field,
such as that produced by stellar bars \citep[e.g.,][]{Spekkens2007,diskfit}, or by a triaxial halo potential. In the presence of an oval distortion particles follows elliptical orbits elongated towards an angle that in general differs from that of the disk position angle \citep[e.g.,][]{Spekkens2007}. 
This kinematic distortion shows a characteristic ``S'' shape in the projected velocity field that makes the minor and major axes not orthogonal \citep[e.g.,][]{Kormendy1983}. Given that this pattern
has been mostly observed in the velocity field of barred galaxies, we will refer to the origin of the oval distortion to stellar bars, although it is not necessarily the case as mentioned before.
The model that intends to describe this pattern is called bisymmetric model \citep[e.g.,][]{Spekkens2007} since most of the perturbation is kept on the  second order of an harmonic decomposition on the disk plane. The bisymmetric model is described by following the expression:
\begin{multline}
 \label{bisymmetric}
 V_\mathrm{bis,model}  = V_\mathrm{sys} + \sin i \Big( V_t(r)\cos \theta - V_{2,t}(r)\cos 2\theta_\mathrm{bar} \cos \theta \\ - V_{2,r}(r) \sin 2\theta_\mathrm{bar} \sin \theta \Big)
\end{multline}
$V_{2,t}$ and $V_{2,r}$ are the nonaxisymmetric velocities induced by the oval distortion and represent, respectively, the tangential and radial deviations from $V_t$, where the latter describes the disk circular rotation.  
The angular variable $\theta_\mathrm{bar}$ is the location relative to the position angle of the bar (\PAbar), in this way:\footnote{Note that the problem becomes degenerated when the bar position angle is aligned to the galaxy major axis. When \PAbar~$ = 0^{\circ}$, the terms $\cos 2(\theta-\phi_{\mathrm{bar}}) \cos\theta$ and $\sin 2(\theta-\phi_{\mathrm{bar}}) \sin \theta$ can be expressed as 
$\frac{1}{2}\big( \cos \theta + \cos 3\theta$ \big) and$ \frac{1}{2}\big( \cos \theta - \cos 3\theta$ \big), respectively. A similar relation occurs when the bar is oriented along the minor axis, \PAbar~$=90^{\circ}$. }:
\begin{equation}
\theta_\mathrm{bar} = \theta -\phi_\mathrm{bar}
\end{equation}
Note that in this expression both angles are measured on the disk plane. If \PAbar~represents the major (minor)--axis position angle of a bar, then both $V_{2,t}(r)$ and $V_{2,r}(r)$ have positive (negative) values.
Unlike the disk position angle \PAdisk, \PAbar~is not a variable than can be easily recognized from the velocity field of barred galaxies; however, its projection in the sky plane is related with \PAdisk~ and the disk inclination angle, as follows:
\begin{equation}
\label{eq:bar_projected}
\phi_\mathrm{bar}^{\prime} =  \phi_{disk}^{\prime} +\arctan(\tan \phi_\mathrm{bar} \cos i)
\end{equation}
where $\phi_\mathrm{bar}^{\prime}$ is the position angle of the bar in the sky plane. Although, computationally it is more practical to estimate $\phi_\mathrm{bar}$ instead of $\phi_\mathrm{bar}^{\prime}$.
In case that the oval distortion is produced by a stellar bar, $\phi_\mathrm{bar}^{\prime}$ is expected to be aligned with the photometric position angle of the  bar, while the radial profile of      $V_{2r}(r)$ and $V_{2t}(r)$ should extend to the length of the bar.
 %Figure~\ref{fig:cartoon} shows an schematization of the variables in the galaxy plane.

\subsection{Harmonic decomposition}
\label{sec:harmonic}
Similar to the photometric decomposition of galaxy images into light profiles via Fourier expansions, the line of sight (LoS) velocity field of a galaxy can be expressed as a sum of harmonic terms as follows:
\begin{equation}
\label{eq:harmonic}
 \mathrm{V_\mathrm{hrm,model}} = V_\mathrm{sys} +  \sum_{m=1}^{M} (c_m(r)\cos m\theta + s_m(r)\sin m\theta) \sin i
\end{equation}
where $c_m$ and $s_m$ are the harmonic velocities,  $m$ is the harmonic number, and $\theta$ and $r$ have the same meaning as  before. For convenience we have taken the inclination angle out of the Fourier expansion; also note that the 0$^\mathrm{th}$ order of the expansion $c_0(r)$ is assumed a constant value equal to the systemic velocity.
However, in addition to the expansion up to $M=1$, where we recover the radial model, note that $c_1 \sim V_t$ and $s_1 \sim V_{r}$; the expansion to higher orders do not offer a direct interpretation of $c_m$ and $s_m$ since these terms represent a mere decomposition of the LoS velocities.
Even though it is possible to assign these velocities a physical meaning. The harmonic number is closely related with perturbations to the gravitational potential; under the epicycle theory, such perturbations will induce the appearance of harmonic sectors in the LoS velocities in such a way that if the gravitational potential contains a perturbation of order $m$, the LoS velocities contain the $m+1$ and $m-1$ harmonic terms of the Fourier expansion \citep[see][for a detailed description]{Schoenmakers1997}. For instance, a bar-potential can be described by a $2^\mathrm{nd}$ order perturbation, which means that the LoS velocity field will contain the $1^\mathrm{st}$ and $3^\mathrm{rd}$ harmonic terms of equation~\ref{eq:harmonic} \citep[e.g.,][]{Wong2004, Kambiz2005}. Similarly, this analysis can be extend for the case of spiral arms \citep[e.g.,][]{vandeVen2010}. 

In \xs~the harmonic model, (equation~\ref{eq:harmonic}),  can be expanded to any harmonic order. Although, most of the non-circular motions induced by spiral arms or bars are captured by a third order expansion \citep[e.g.,][]{Trachternach2008}.

The harmonic model was first included in  the GIPSY task {\tt RESWRI} \citep[e.g.,][]{RESWRI} under the assumption of thin disk. Afterwards the harmonic decomposition was generalized in {\tt KINEMETRY} \citep[e.g.,][]{Kinemetry}  including not only disks but also triaxial structures.
%
%As we will see in forthcoming sections, \xs~allows fitting for any harmonic number ({\tt hrm\_m})\footnote{From here and on, all the \xs~inputs will be shown in typewriter font.}, without the need of making a previous circular model. At difference from {\tt KINEMETRY} and {\tt RESWRI}, \xs~fits the entire velocity field in one single fit. Thus it takes information from all pixels to derive the best two dimensional model. 
%
The major difference between {\tt RESWRI} and \xs~is the assumption of a flat disk. 
While {\tt RESWRI} and {\tt KINEMETRY} allow varying the disk position angle and inclination during the fitting analysis, 
\xs~keeps these angles fixed to allow the residual velocities of a circular model to be adjusted with non-circular motions and not absorbed by the variations of these angles, as explained before.
However, when large variations of \PAdisk~or $i$ are present throughout the disk, \xs~will fail in the interpretation of the harmonic velocities, even when the fit is successful.

%The harmonic number $m$ in Equation~\ref{harmonic} is related with the gravitational potential. Perturbations in the potential of order $m$ affect only the $(m+1)$ and $(m-1)$ coefficients.

\section{the algorithm}
\xs~works on 2D velocity maps, such as those extracted from first moment maps from datacubes.
As others codes that rely on 2D maps for kinematic modelling, \xs~assumes that the velocity recorded in each pixel is representative of the disk
velocity. In this sense, there are a wide variety of methods for representing the velocity field of a galaxy, and much of these depend on the spectral resolution and
the signal-to-noise of the data;
going from simple first moment maps, to modeling  Gaussian profiles in combination with Hermite polynomials to better reproduce the
shape of the emission lines \citep[see][for a revision of different methods]{THINGS, Sellwood2021}.

For \xs~to obtain confident estimations of the kinematic models, the data should not be strongly affected by the point-spread-function. The PSF contributes to increase the velocity dispersion of the emission-lines and consequently to underestimate the rotation velocities; particularly in the inner gradient of the rotation curve.
In such a case, a 3D modeling of the datacubes should be a better approach \citep[e.g.,][]{DiTeodoro2016}.
%
%In addition, \xs~is intended to work on velocity maps not strongly affected by spatial resolution effects (i.e., beam smearing).
%We note that including a 2D Gaussian convolution on the models during the minimization process slow down the fitting by various orders of magnitude;
%despite of that, this function is included, although it is highly inefficient.
%
%However, low angular resolution data will be more affected by the point-spread-function (PSF), particularly in the inner gradient of the rotation curve. In such cases, a previously PSF corrected velocity map should be passed, for instance \citet{Chung2020}.
%
Under the previous assumptions the algorithm proceeds in the following way.

Let $(x_n,y_n)$ be the position of a data point in the sky plane. The corresponding ellipse passing through this point, with center $(x_0,~y_0)$ and rotated by an angle $\phi_\mathrm{disk}^\prime$ is described by:
\begin{align}
x_e & = -(-x_n-x_0)\sin \phi_\mathrm{disk}^\prime + (y_n-y_0)\cos \phi_\mathrm{disk}^\prime\\
y_e & = -(x_n-x_0)\cos \phi_\mathrm{disk}^\prime -(y_n-y_0)\sin \phi_\mathrm{disk}^\prime 
\end{align}
The radius of the circle on the disk plane passing through this point is then:
\begin{equation}
 r_n^2 = x_e^2 + \Big( \frac{y_e}{\cos i} \Big)^2
 \end{equation}
 The azimuthal angle on the disk plane $\theta$, is related to the sky coordinates as follows:
\begin{align}
\cos \theta & = \frac{ -(-x_n-x_0)\sin \phi_\mathrm{disk}^\prime + (y_n-y_0)\cos \phi_\mathrm{disk}^\prime}{r_n}\\
\sin \theta & = \frac{ -(x_n-x_0)\cos \phi_\mathrm{disk}^\prime -(y_n-y_0)\sin \phi_\mathrm{disk}^\prime}{r_n \cos i} 
\end{align}

Therefore, $\theta$ comprises both projection angles \PAdisk~and $i$, as well as the kinematic center, thus contributing with four more free variables in each kinematic model (although it is represented by a single variable for simplicity). 
Henceforth, we define ``constant parameters'' as those variables that do not change with radius, these are:  $\langle \phi_\mathrm{disk}^\prime,~i,~x_0,~y_0,~V_{sys},~\phi_\mathrm{bar} \rangle$. We will also refer to ``geometric parameters'' to those variables that describe the orientation of the projected ellipse on the sky plane, namely $\phi_\mathrm{disk}^\prime,~i,~x_0,~y_0$.  

\subsection{$\chi^2$ minimization technique}
As we will see in further sections, Bayesian methods like MCMC require start sampling around the {\it maximum a posteriori}, or maximum likelihood, to generate new samples also known as chains; this requires necessarily to find those parameters that minimize the residuals from a given kinematic model and the data.
Therefore, in the following we describe the method to solve for each of the different kinematic components of the models, and the constant parameters. 
The first part corresponds to the analysis adopted in {\tt DiskFit} \citep[e.g.,][]{Spekkens2007,diskfit}, with minimum changes.

A given set of initial conditions for the geometric parameters defines the projected disk with an elliptical shape  on the  sky plane.  Ideally, the initial conditions for the gaseous disk geometry should be close to that of the stellar disk. This geometry will be the starting configuration for the minimization analysis; then, the field of view is divided into $K$ concentric rings of fixed width that follow the same orientation as before. The maximum length of ellipse semi-major axis can be easily set-up as described in the Appendix~\ref{sec:appendix}; this will create a 2D mask and only those pixels inside this maximum ellipse will be considered for the analysis. The geometry of this mask will be adapted in subsequent iterations until reaching the orientation that better describes the observed velocity field.

%Once defined the starting geometry of the projected disk,  we can proceed to solve for the different velocity components included in Eqs.~\ref{circular}--\ref{bisymmetric} and Eq.~\ref{eq:harmonic}. %This procedure is performed by dividing the velocity field in $K$ concentric rings oriented with the same constant parameters as the projected disk. The width of the rings is constant and it is defined as {\tt 2delta}, where delta is half of the ring width.
%
The algorithm will solve for each ring, a set of velocities that will depend on the kinematic model considered, namely Eqs.~\ref{circular}--\ref{bisymmetric} or Eq.~\ref{eq:harmonic} . Thus, the number of different velocity components to derive will be $K$  velocities in the circular model ($V_{t,K}$); $2K$ in the radial model ($V_{t,K},~V_{r,K}$); $3K$ in the bisymmetric model ($V_{t,K},~V_{2r,K},~V_{2t,K}$) and $2MK$ in the harmonic model ($ c_{1,K},...,c_{M,K}, s_{1,K},...,s_{M,K}$).

The velocity map consists of a two--dimensional image of size $nx \times ny$, with 
$N$ observed data points $\mathcal{D}_n$ with individual errors $\sigma_n$. 
Let $\overrightarrow V$ be the set of velocities that describe the corresponding kinematic model (namely, $\overrightarrow V =\langle \overrightarrow  V_t,~ \overrightarrow V_{2,t}, \overrightarrow V_{2,r} \rangle$ for the bisymmetric model and similarly for other models). 
Frequentist methods adopt the chi-square $\chi^2$, to derive from a model the set of parameters that describes the data.
In this case, the reduced $\chi_r^2$ for the different kinematic models is given by:
\begin{equation}
 \label{Eq:chi2}
 \chi_r^2 = \frac{1}{\nu} \sum_{n = 1}^N \Big(\frac{\mathcal{D}_n - \sum_{k=1}^K W_{k,n}\overrightarrow V_k }{\sigma_n} \Big)^2 
\end{equation}
Here $\nu$ is the total number of degrees of freedom (i.e., $\nu = N - N_\mathrm{varys}$, and $N_\mathrm{varys}$ is the number of parameters to estimate from the model );  $W_{k,n}$ are a set of weights that depend on the pixel position, and will serve to define an interpolated model; and $\overrightarrow V_k$ is the set of velocities in the $k$-th ring that describes the considered kinematic model.

Each kinematic component from $\overrightarrow V_k$ would require different weights. For instance, for the circular model the weights adopt the following expression:
\begin{equation}
\label{w_circ}
 W_{k,n}^t = \sin i \cos \theta*w_{k,n}
\end{equation}
where the super-index $t$ makes reference to the circular rotation component. The radial model requires two different weights for the different kinematic components:
\begin{align}
\label{w_rad}
\centering
 W_{k,n}^t  & = \sin i \cos \theta*w_{k,n}\\
W_{k,n}^r   & = \sin i \sin \theta*w_{k,n}
\end{align}
Similarly, the bisymmetric model would require three different weights:
\begin{align}
\label{w_bissym}
\centering
 W_{k,n}^t  & = \sin i \cos \theta*w_{k,n}\\
W_{k,n}^{2r}   & = \sin i \cos \theta \cos 2\theta_\mathrm{bar}*w_{k,n}\\
W_{k,n}^{2t}   & = \sin i \sin \theta \sin 2\theta_\mathrm{bar}*w_{k,n}
\end{align}
Finally, the harmonic decomposition model will have $2M$ weights given by:
\begin{align}
\label{w_harm}
\centering
W_{k,n}^c  & = \sum_{m=1}^{M} \sin i \cos m\theta*w_{k,n}\\
W_{k,n}^s   & = \sum_{m=1}^{M} \sin i \sin m\theta*w_{k,n}
\end{align}
Note that for $M=1$ it reduces to the radial model.

The $w_{k,n}$ terms define the interpolation method to be performed between the $K$--rings. As in \diskfit, these weights adopt the form of a simple linear interpolation given by the usual expression:
%
%\begin{align}
%\label{w_bisym}
%\centering
%w_{k,n}  & = \big (1 +  \frac{r_n - r_k}{\delta r_k} \big) \\
%w_{k+1,n}  & = \big ( \frac{r_n - r_k}{\delta r_k} \big)
%\end{align}

\begin{align}
\label{eq:weights_interp}
\centering
w_{k,n}  & = \bigg (\frac{r_{k+1} - r_n}{\delta r_k} \bigg) \\
w_{k+1,n}  & = \bigg ( \frac{r_n - r_k}{\delta r_k} \bigg)
\end{align}

where $r_k$ and $r_{k+1}$ are the position of the $k^{th}$ and $(k+1)^{th}$ rings respectively, and $\delta r_{k} = r_{k+1} - r_{k}$ is the spacing between rings. 
As the first ring $(k = 1)$ can not be placed at the kinematic centre, \xs~
implements different strategies for assigning velocities to pixels down the first ring. Depending on the spatial resolution of the data or the signal-to-noise ratio (\SN), one may opt for one of the following extrapolation options. 

The first method is to assume that velocities grow linearly from zero to the velocities derived in the first ring ($ \overrightarrow V_1$). This implies that $\overrightarrow V_0 =0$ at $r = 0$; therefore, the kinematic center does not rotate.
In the second approach, the set of velocities and positions $\big( \overrightarrow V_1, r_1 \big)$ and $\big( \overrightarrow V_2, r_2 \big)$ are used to extrapolate velocities to pixels down $r_1$; in this way $\overrightarrow V_0 \neq0$ at $r = 0$.
The third option allows the user to fix the velocity at the origin to some value. Then $\overrightarrow V_0$ and $\overrightarrow V_1$ are linearly interpolated for sampling pixels down $r_1$.

As $\overrightarrow V_k$ is linear in Eqs.\ref{circular}--\ref{bisymmetric} and Eq.~\ref{eq:harmonic}, we can set the derivative with respect to $\overrightarrow V_j$ in Eq.~\ref{Eq:chi2}, giving as result:
\begin{equation}
 \label{derivate_1}
 \frac{\partial \chi_r^2}{\partial \overrightarrow V_j} = \frac{2}{\nu} \sum_{n = 1}^N \Big(  \frac{\mathcal{D}_n - \sum_{k=1}^K W_{k,n}\overrightarrow V_k }{\sigma_n} \Big) \frac{W_{j,n}}{\sigma_n} = 0
\end{equation}
Rearranging this expression we obtain:
\begin{equation}
 \label{derivate_2}
 \sum_{k = 1}^K \Big( \sum_{n=1}^N \frac{W_{k,n}}{\sigma_n} \frac{W_{j,n}}{\sigma_n} \Big) \overrightarrow V_k = \sum_{n = 1}^N \frac{W_{j,n}}{\sigma_n^2} \mathcal{D}_n
\end{equation}
The minimization technique from Eq.~\ref{derivate_2} was first introduced by \citet{Barnes2003}, and subsequently incorporated into \diskfit~
\citep{diskfit}. Here, Eq.~\ref{derivate_2} is generalized for the harmonic decomposition model. 
The latter expression is a system of linear equations for the $\overrightarrow V_k$ unknowns; thus $\overrightarrow V_k$ values are solved arithmetically.
As mentioned before, the number of velocity components $\overrightarrow V_k$ depends on the number of rings and the adopted kinematic model, thereby the dimensions of the matrix to solve will increase as more rings are included in the analysis, and as the kinematic model becomes more complex.

%where $V_k$ is $(V_t, V_{2,t}, V_{2,r})$ for the bisymmetric model;  $(V_t, V_{r})$ for the radial model; or $V_t$ for the circular model.

Given a set values for the constant parameters and  $K$ rings positioned at $r_k$ on the disk plane, we can solve for $\overrightarrow V_k$ in Equation~\ref{derivate_2} by assigning
uniform weighting factors ($w_{k,n}=1$). This way we are obtaining a row-stacked velocities ($r_k$ vs. $V_k$). In essence the row-stacked velocities represent the average velocity of each ring; then, we use these velocities as initial conditions to perform an iteratively least squares analysis (LS) through Equation~\ref{Eq:chi2}, but now with the proper weighting factors (namely Eqs.~\ref{w_circ}--\ref{w_harm}). Note that if the initial geometric parameters passed to the algorithm are close the true ones, then the arithmetic solutions to $\overrightarrow V_k$ should be close the true velocities. This can speedup the MCMC sampling as we will see in further sections.

\begin{figure*}[t!]
\centering
\includegraphics[]{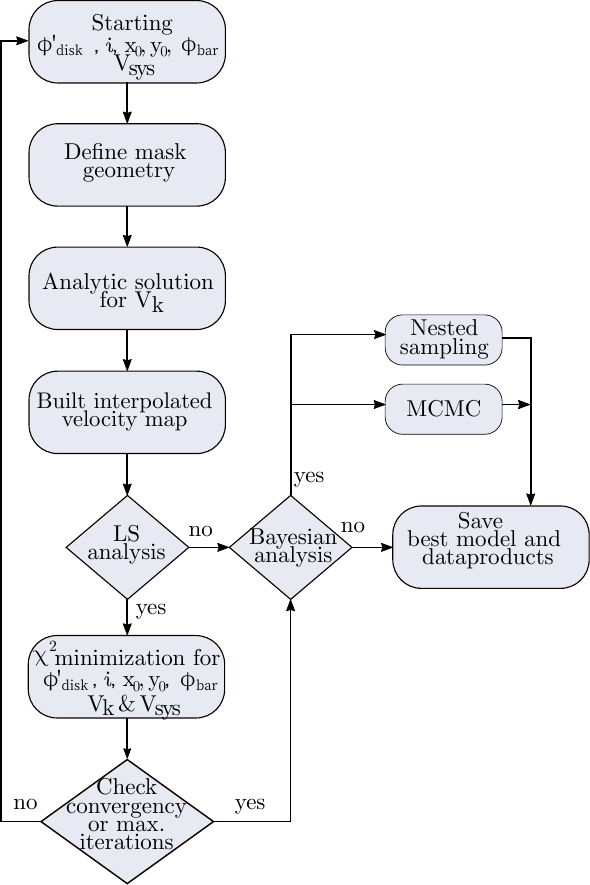}
\caption{Flowchart of the fitting procedure to derive the best kinematic model.}
\label{fig:flowchart}
\end{figure*}

The minimization procedure in Equation~\ref{Eq:chi2} is performed by constructing a 2D model from the interpolation weights of Eq~\ref{eq:weights_interp}.
In each $\chi^2$ iteration a new set of velocities $\overrightarrow V_k^{\prime}$ are obtained, together with a new set of constant parameters. The latter will define a new geometry for the mask, and new row-stacked velocities will be obtained for the next iteration. Multiple rounds of $\chi^2$ minimization will be performed up to some maximum iteration defined by the user, or until the difference in $\chi^2$ evaluations varies less that $10\%$. Commonly after three  iterations the disk geometry becomes stable and simultaneously $\overrightarrow V_K$. Figure~\ref{fig:flowchart} sumarizes all the fitting procedure in a flowchart. 

Rings not proper sampled with data may give absurd values of $\overrightarrow V_k$ when solving Eq.~\ref{derivate_2}. To avoid this problem, we define a covering factor to guarantee a minimum number of data per ring. If the covering factor is 1, it means that rings must be 100\% occupied by data to estimate $\overrightarrow V_k$, as described in Appendix~\ref{sec:appendix}. 
In addition, isolated pixels in the image  due to low \SN~may not be desired during the analysis; \xs~allows to remove these pixels by excluding those with velocity errors greater than certain 
threshold defined by the user.

For performing the LS analysis in equation~\ref{Eq:chi2}, \xs~adopts the Levenberg--Marquardt (LM) algorithm included in the {\tt lmfit} package \citep[e.g.,][]{lmfit}. This algorithm has the advantage that is fast, although it is widely known to be susceptible to getting trapped at a local minimum.
Note that \texttt{DiskFit} adopts the Powell method since this method only performs evaluation  of functions with no derivatives performed.

So far the algorithm adopts a LS method for deriving the best parameters defined by the kinematic models.
%Angular variables such as the position angles may give rise to multi-modal distributions; thus, it is important to sample all parameter space to avoid being trapped in local minimums of the $\chi^2$.
%
In the following we use sampling methods to infer the posterior distributions of the parameters.

\subsection{Bayesian analysis}

The novelty of \xs~resides on the estimation of the posterior distribution of the non--circular motions and the model parameters. Given the high-dimension of the models, it is desired to perform a thorough analysis of the prior space to obtain the most likely solutions to the problem for each kinematic model regardless of its complexity.
For this purpose we adopt Bayesian inference methods for sampling their posterior distributions. 
Other packages like {\sc 2dbat} and {\sc KinMSpy} \citep[i.e.,][]{2DBAT, kinms} also use Bayesian approaches for extracting rotational curves of galaxies. The difference is that {\sc KinMSpy} is able to fit non--circular motions (radial and bisymmetric).

%According to the Bayes' theorem, for a given set of data $\mathcal{D}$, the posterior distribution $p$ of the parameters $\vec {\alpha} = \alpha_1,~\alpha_2,~..,~\alpha_n$ is given by :
According to Bayes' theorem, 
given a set of data $\mathcal{D}$ described by a model function $\mathcal{M}$ with parameters $\vec {\alpha} = \alpha_1,~\alpha_2,~..,~\alpha_n$, the posterior distribution of $\vec {\alpha}$ given $\mathcal{D}$ follows the expression: 

\begin{equation}
\label{Eq:Bayes}
 p(\vec {\alpha}|\mathcal{D},\mathcal{M}) = \frac{p(\mathcal{D}|\vec {\alpha},\mathcal{M}) p(\vec {\alpha})}{p(\mathcal{D},\mathcal{M})}
\end{equation}
where $p(\vec {\alpha}|\mathcal{D},\mathcal{M})$ is the joint posterior distribution of the whole set of parameters; $p(\mathcal{D}|\vec {\alpha},\mathcal{M})$ is the probability density of the data given the parameters and the assumed model; $p(\vec {\alpha})$ is the prior probability distribution of the parameters and
$p(\mathcal{D},\mathcal{M})$ is a normalization constant also know as {\it marginal evidence} or evidence.
It is common to find Eq.~\ref{Eq:Bayes} expressed in terms of the likelihood function $\mathcal{L}$, as follows:
\begin{equation}
%\large
p(\vec {\alpha}|\mathcal{D},\mathcal{M}) = \frac{\mathcal{L}(\vec {\alpha}) p(\vec {\alpha})}{\mathcal{Z}}
\end{equation}
with the evidence defined as:
\begin{equation}
\label{Eq:Evidence}
%\large
\mathcal{Z}=\int_{\Omega_{\alpha}} \mathcal{L}(\vec {\alpha}) p(\vec {\alpha}) d\vec {\alpha}
\end{equation}
where the integral is computed over all the parameter space defined by the priors, $\Omega_{\alpha}$. The evidence can be interpreted as the likelihood of the observed data under the model assumptions; in other words, it is the average of the likelihood over the priors.

The final goal of Bayesian inference is to obtain the posterior distribution $p(\vec {\alpha}|\mathcal{D},\mathcal{M})$ of all  parameters $\vec {\alpha}$ describing the model function $\mathcal{M}$.
Multiple methods  have been developed for this purpose. For instance, Markov-Chain Monte Carlo (MCMC) methods evaluate the unnormalized posterior distribution (i.e., $p(\vec {\alpha}|\mathcal{D},\mathcal{M}) \propto \mathcal{L}(\vec {\alpha}) p(\vec {\alpha})$), by generating samples or chains from the likelihood function. One of the main characteristics of Markov chains is that the position of a point in the chain depends only on the position of the previous step.
Different algorithms with automating chain tunning have been developed to efficiently sample the posterior distribution. Among the most popular MCMC samplers are those who implement {\it affine-invariant ensemble sampling} and {\it ensamble slice sampling} \citep[e.g.,][]{emcee, karamanis2021zeus}.

Other methods, such as nested sampling  \citep[NS, ][]{nested}, are designed to
compute the evidence by  numerical integration of Eq.~\ref{Eq:Evidence}, which often makes them computationally more expensive than MCMC methods. 
This integral is performed from the priors space (or prior volume), and unlike MCMC, does not require an initialization point.
Nevertheless, computing the evidence is crucial for model comparison as it represents the degree to which the data is in agreement with the model. Although the main goal of NS is to compute the evidence, the posterior distribution is obtained as a by-product; because of that, NS methods are becoming popular for the inference of parameters in astronomy \citep[see][for a thorough description of the method]{Ashton2022}. 

One of the advantages of nested sampling with respect MCMC methods is regarding the convergence criteria. 
 There is no defined convergence criteria among MCMC algorithms, although some of them
 are based on the number of independent samples in the chains, the so-called
 { \it itegrated autocorrelation time (IAT)}; however this is often evaluated a posteriori. If the whole chain contains between 10-50 times the IAT, then it is a good indicator that chains are converging \citep{emcee,karamanis2021zeus}.
In contrast, in nested sampling the stopping evaluation criterion is well defined, since sampling stops after the whole prior space has been integrated.

A detailed discussion of these two sampling methods is however, beyond the scope of this paper. Following we show the implementation of MCMC and nested sampling methods for the parameter extraction of the kinematic models presented in Sec.~\ref{sec:kinematic_models}.

%For sampling the posterior distribution we use Markov-Chain Monte Carlo (MCMC) methods. Advantages of using MCMC is the ability of sampling the posterior distribution regardless of the number of dimension of the problem .

\subsubsection{Likelihood and priors}

Let $\vec{\alpha}$ be all the parameters that describe any of the kinematic models. Then, the log posterior distribution of the parameters is given by:
\begin{equation}
\label{bayes}
\ln p(\vec{\alpha} | \mathcal{D},\mathcal{M}) = \ln \mathcal{L}(\vec {\alpha})+  \ln p(\vec{\alpha}) - \ln \mathcal{Z}
\end{equation}

The likelihood function is a key term in Bayesian inference, since it will define the shape of the posterior distributions. The most common distribution for the likelihood is Gaussian, but other distributions like Cauchy, T-student, or the absolute value of the residuals are also adopted in the literature \citep[e.g.,][]{3Dbarolo, Galpack, 2DBAT}. \xs~adopts the Gaussian distribution as the main likelihood function, although Cauchy distribution is also included (see Appendix~\ref{sec:cauchy}). The individual likelihood for each data point $\mathcal{D}_n$ with error $\sigma_n$ is expressed as:
\begin{equation}
 \mathcal{L}(\alpha_n) = (2\pi\sigma_n)^{-1/2}~ \mathrm{exp}(-\frac{(\mathcal{D}_n-\mathcal{M}_n)^2}{2\sigma_n^2})  
\end{equation}
and the joint likelihood for the data set is the product of individual likelihoods, in this way
\begin{equation}
 \mathcal{L} = (2\pi)^{-N/2}~ \Big( \prod_{n=1}^N \sigma_n \Big) \mathrm{exp}(- \sum_{n=1}^N \frac{(\mathcal{D}_n-\mathcal{M}_n)^2}{2\sigma_n^2})  
\end{equation}
It is easy to recognize from this expression that the summation is the $\chi^2$ from Eq~\ref{Eq:chi2}, with $\mathcal{M}_n$ being the kinematic model function, $\mathcal{V}_\mathrm{model}$.
In this way the log posterior distribution of the parameters is expressed as:
\begin{multline}
\label{eq:logpost}
  \ln p(\vec{\alpha} | \mathcal{D},\mathcal{V}_\mathrm{model}) =  -\frac{1}{2} \sum_{n=1}^N  \frac{ \big(\mathcal{D}_n - \sum_{k=1}^K W_{k,n} \overrightarrow V_k \big)^2 }{\sigma_n^2}  \\
          -\ln \sigma_n - \frac{N}{2} \ln(2\pi)  + \ln p(\alpha) - \ln \mathcal{Z}
\end{multline}
with $N$ being the number of data points, or pixels, to be considered in the model.
We can redefine $\sigma$ to include the intrinsic dispersion of the data, which we assume constant for all pixels; namely, $\sigma_n^2 = \sigma_n^2 + \sigma_{int}^2$.

\begin{table}\centering
\setlength{\tabnotewidth}{1\columnwidth}
\tablecols{3}
\setlength{\tabcolsep}{2.8\tabcolsep}
\caption{Type of priors adopted in \xs.}
\begin{tabular}{lll}
\hline
{Parameter} & {Uniform prior}& {Truncated Gaussians\tabnotemark{a}} \\
\hline
$\phi_{\mathrm{disk}}$ & 0 if $-2\pi<\phi_{\mathrm{disk}}<2\pi$ & TG($\widehat{\phi}_{\mathrm{disk}}$,$15^{\circ}$,$\mp45$) \\
$i$ & 0 if $30<i<75$ & TG($\widehat{i}$,$10^{\circ}$,30,75) \\
$\phi_{\mathrm{bar}}$ & 0 if $-\pi<\phi_{\mathrm{bar}}<\pi$ & TG($\widehat{\phi}_{\mathrm{bar}}$,$20^{\circ}$,$\widehat{\phi}_{\mathrm{bar}}\mp45$) \\
$x_0$ & 0 if $0<x_0<nx$ & TG($\widehat{x_0}$,$2\arcsec$,$\widehat{x_0}\mp10\arcsec$) \\
$y_0$ & 0 if $0<y_0<ny$ & TG($\widehat{y_0}$,$2\arcsec$,$\widehat{y_0}\mp10\arcsec$) \\
$V_{sys}$ & 0 & TG($\widehat{V}_{sys}$,50~\kms) \\
$V_{k}$ & 0 if $-400<V_k<400$ & TG($\widehat{V}_{k}$,150~\kms,-250,250) \\
$\ln \sigma^2_{int}$ & 0 if $-10<\ln \sigma^2_{int}<10$ & TG(0.1,1,-10,10) \\
\hline
\tabnotetext{a}{\small Values with hat represent LS results.}
\tabnotetext{}{\small$V_k$ refers to any of the different radial dependent velocities.}
\end{tabular}
\label{Tab:priors}
\end{table}
The priors are the constrain of our model function and enclose all we know about the data. Uniform or non-informative priors give the same probability to any point within
the considered boundaries. This allows the likelihood function to survey the prior space without any preferred direction.
\xs~adopts either uniform or truncated Gaussians (TG), with values shown in Table~\ref{Tab:priors}. TG priors are of the form TG($\mu$,$\sigma$,$\mu_{min}$,$\mu_{max}$), with $\mu$ and $\sigma$ being the mean and standard deviation of the Gaussian, and $\mu_{min}$ and $\mu_{max}$ represent the lower and upper boundaries respectively. The mean values can be chosen arbitrary, although good values are those that maximize the likelihood function (i.e., Eq.~\ref{Eq:chi2}). In most cases, choosing uniform or TG priors does not affect the posterior distributions. The difference resides in the computational cost needed to explore the prior space; narrow distributions like TG are sampled more efficiently rather than uniform distributions.

In order to infer the posterior distribution of the parameters, \xs~ adopts two well known
Python packages for Bayesian analysis; these are the {\sc emcee} package \citep[e.g.,][]{emcee}, and {\sc dynesty}  \citep[e.g.,][]{dynesty}. {\sc emcee} is a Python implementation of the affine-invariant method for MCMC with automatic chain tuning; while {\sc dynesty} is a Python implementation of dynamic nested sampling  methods. 
MCMC and NS are two robust sampling techniques to derive posterior distributions in high dimensional likelihood functions, such as the kinematic models described before.
Both packages have been extensively applied in astronomy for making Bayesian inference, with particular implementations in cosmology.
For a detail description of these codes we suggest reading their corresponding documentation. 
Both packages require a set of configurations that have for purpose guarantee convergence of the sampling procedure.
\xs~is optimized to pass a configuration file to set up {\sc emcee} and {\sc dynesty}. The main setups in these codes
are the length of the join-chains and the discarding fraction (burning period) in the case of MCMC, and the integration limit for NS.
For both packages, \xs~adapts the likelihood functions and priors to make it compatible with MCMC or NS methods.

As mentioned before, MCMC samplers like {\sc emcee} sample from the likelihood; therefore 
the chains need to be initialized at some position, for which \xs~ chooses a random region around the maximum likelihood.
For MCMC samplers the joint posterior distribution is estimated up to a normalization constant, here adopted equal to 1 (or zero in ln). 
For running {\sc dynesty} \xs~ transform the priors from Table~\ref{Tab:priors} into a unit cube, in such a way that all parameters vary from 0 to 1 and they are re-scaled at the end of the sampling process.  

Finally, representative values of the parameters are taken as the 50\%~percentile of the marginalized distributions. The uncertainty on the parameters is addressed in the following section.

Although {\sc emcee} makes use of frequentist methods for starting the sampling process, this could be suppressed if relatively good initial positions of the disk geometry are 
given. On the other hand, {\sc dynesty} does not require at all the LS initialization as the numerical integration is performed over the prior space.

\subsubsection{Error estimation}
\label{sec:best_fit_error}
The true uncertainty in rotational velocities are known to be underestimated with standard least squares minimization techniques and even with MCMC methods \citep[e.g.,][]{THINGS,2DBAT}. Errors estimated with these methods are usually  of the order of the turbulence of the ISM (a few \kms) and do not represent the systematic errors. Some works adopt the mean dispersion per ring as a measure of the uncertainty in the rotation curve. However, when non--circular components are added to the model, this assumption is no longer valid since each ring may contain multiple kinematic components.

\xs~provides different error estimates on the derived parameters. The Levenberg--Marquardt least-squares minimization automatically computes errors from the covariance matrix; these represent statistical errors and may be used for a quick analysis.  
However, the power of Bayesian inference relies on the estimation of posterior distributions, from which we can obtain uncertainties on the parameters.
\xs~adopts the marginalized distributions to quote the uncertainties in each parameter, including the velocities. 
These uncertainties are in general smaller than simple Monte-Carlo errors since  marginalized distributions are not expected to contain unstable {\it(burn-in)} chains. This necessarily requires dropping an important fraction of the total samples during a run, which is customized within \xs.
Therefore, it is common to report $2\sigma$ credible intervals in Bayesian analysis.
In fact, finding ``large'' uncertainties in MCMC methods would be an indication that chains are not fully converging; either because a large fraction of samples are being rejected, or chains are surveying complicated likelihood functions, often multi-modal distributions, for which NS would be a better solution. 

Additionally, \xs~also implements a bootstrap analysis for error estimation. Our procedure differs from the one described by \citet{Sellwood2010}, as explained below.
The residuals from the best 2D interpolated model are used to generate new samples. Instead of shuffling residuals at random locations on the disk, $K$ rings of width $\delta_r$ are constructed with projection angles given by the best values. Then, residuals in each ring are chosen to resample the best 2D model in the same ring locations. In this way any residual pattern associated to a bar or spiral arms remains around the same galactocentric distance but not in the same pixel location. The new re-sampled velocity map is used in a least squares analysis for deriving a new set of velocities and constant parameters.
This procedure is performed iteratively; finally the root mean square deviation  is  taken as $1\sigma$ error;  however, for consistence with the Bayesian methods, we report $2\sigma$ errors throughout the paper. 

In general, we find that the estimated uncertainties on the parameters increase in the following order: Bayesian methods $>$ bootstraps $>$ LS, with computational cost increasing in the same direction.

\begin{table}[t]
\setlength{\tabnotewidth}{1\columnwidth}
\tablecols{10}
\small
\begin{changemargin}{-2cm}{-2cm}
\caption{Best fit parameters for the toy model example.}
\label{tab:res_simulation}
\begin{tabular}{llcccccccc}
    \hline
    {Model} & method  & {$\Delta$\PAdisk} & {$\Delta i$} & {$\Delta x_0$}& {$\Delta y_0$} &  {$\Delta V_\mathrm{sys}$} & {$\Delta$\PAbar} & RMS & BIC \\ 
            &         & $(^\circ)$ & $(^\circ)$ & $(pix)$ & $(pix)$ &  (\kms) & $(^\circ)$ & (\kms) &  \\ 
    \hline
    circular & LM & 
    -1.1 $\pm$ 0.1 & -0.4 $\pm$ 0.1 & 0.0 $\pm$ 0.0 & 0.1 $\pm$ 0.0 & -0.0 $\pm$ 0.2 & \nodata & 8.5 & 4.5 \\
    &MCMC&-1.1 $\pm$ 0.1 & -0.1 $\pm$ 0.3 & 0.0 $\pm$ 0.1 & 0.1 $\pm$ 0.1 & -0.0 $\pm$ 0.2 & \nodata & 9.3 & 4.5 \\
    &NS&-1.1 $\pm$ 0.1 & -0.1 $\pm$ 0.3 & 0.0 $\pm$ 0.1 & 0.1 $\pm$ 0.1 & -0.0 $\pm$ 0.2 & \nodata & 9.3 & 4.5 \\
    \hline
    radial & LM &
    -0.1 $\pm$ 0.1 & -0.2 $\pm$ 0.2 & 0.0 $\pm$ 0.0 & 0.1 $\pm$ 0.0 & 0.0 $\pm$ 0.2 & \nodata & 9.3 & 4.4 \\
    &MCMC&-0.1 $\pm$ 0.1 & 0.1 $\pm$ 0.3 & 0.0 $\pm$ 0.1 & 0.1 $\pm$ 0.1 & -0.0 $\pm$ 0.2 & \nodata & 8.5 & 4.4 \\
    &NS&-0.1 $\pm$ 0.1 & 0.0 $\pm$ 0.3 & 0.0 $\pm$ 0.1 & 0.1 $\pm$ 0.1 & -0.0 $\pm$ 0.2 & \nodata & 8.5 & 4.4 \\
    \hline
    bisymmetric & LM &
    0.0 $\pm$ 0.1 & 0.1 $\pm$ 0.2 & 0.0 $\pm$ 0.0 & 0.1 $\pm$ 0.0 & -0.0 $\pm$ 0.2 & 3.8 $\pm$ 8.0 & 8.5 & 4.4 \\
    &MCMC&0.0 $\pm$ 0.2 & 0.1 $\pm$ 0.3 & 0.0 $\pm$ 0.1 & 0.1 $\pm$ 0.1 & -0.0 $\pm$ 0.2 & 3.7 $\pm$ 7.8 & 8.5 & 4.4 \\
    &NS&0.0 $\pm$ 0.2 & 0.1 $\pm$ 0.3 & 0.0 $\pm$ 0.1 & 0.1 $\pm$ 0.1 & -0.0 $\pm$ 0.2 & 3.6 $\pm$ 8.2 & 8.5 & 4.4 \\
    \hline
    harmonic & LM &
    -0.1 $\pm$ 0.1 & -0.0 $\pm$ 0.3 & 0.0 $\pm$ 0.0 & 0.1 $\pm$ 0.0 & -0.0 $\pm$ 0.2 & \nodata & 8.5 & 4.4 \\
    &MCMC&-0.1 $\pm$ 0.1 & 0.0 $\pm$ 0.3 & 0.0 $\pm$ 0.1 & 0.1 $\pm$ 0.1 & -0.0 $\pm$ 0.2 & \nodata & 8.5 & 4.4 \\
    &NS&-0.1 $\pm$ 0.1 & 0.0 $\pm$ 0.3 & 0.0 $\pm$ 0.1 & 0.1 $\pm$ 0.1 & -0.0 $\pm$ 0.2 & \nodata & 8.5 & 4.4 \\
    \hline
    \tabnotetext{}{$\Delta \equiv \alpha_\mathrm{recovered}-\alpha_\mathrm{true}$}
\end{tabular}
\end{changemargin}
\end{table}

\begin{figure*}
\centering\includegraphics[ width = \columnwidth, clip, ]{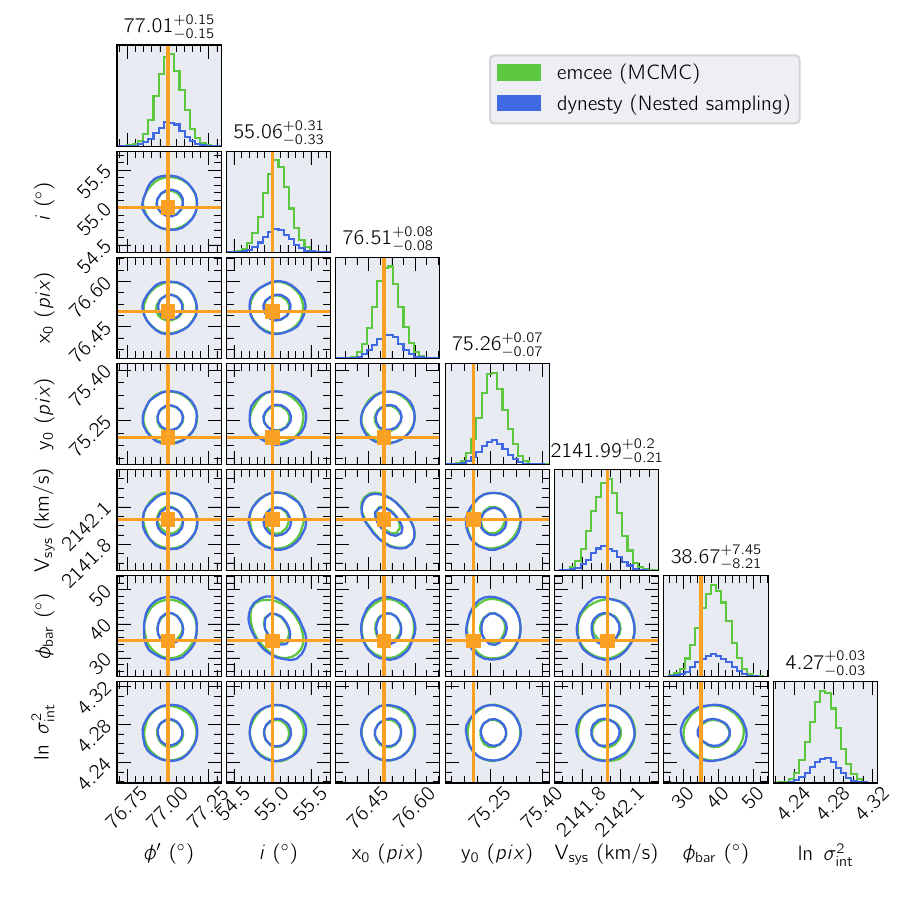}
\caption{Marginalized distributions of the parameters describing the bisymmetric model for our toy model described in Sec.~\ref{sec:single_calse}. This corner plot shows only the parameters describing the disk geometry. Contours enclose the $68\%$ and $95\%$ of the data. 
Histograms of individual distributions are shown on top, together with the median values and $2\sigma$ credible intervals for each parameter. The orange straight lines represent the true values. As observed, all parameters but $y_0$, are recovered within the $\pm1\sigma$ region.  
}
\label{fig:posteriors}
\end{figure*}

\begin{figure*}[t]
\hspace*{-0.1cm}\includegraphics[width = \columnwidth]{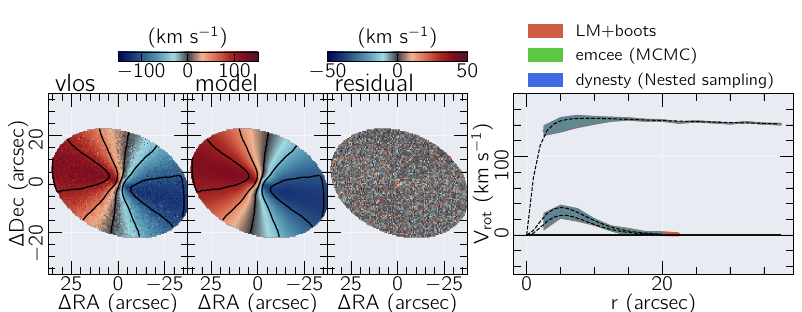}
\caption{\xs~results for the toy model example, for the bisymmetric model case. Figures from left to right: simulated velocity field; best two dimensional interpolated model from MCMC; residual map (input minus output).  Overlaid on these maps are iso--velocity contours starting in 0~\kms~with steps of $\pm50$~\kms.
The fourth column shows the radial variation of the different kinematic components included in the model. Here the input velocities are shown with discontinuous lines while continuous lines represent the velocities derived by \xs~using MCMC (green), nested sampling (blue) and LM + bootstrap (red). The shadowed regions represent $2\sigma$ errors.}
\label{fig:toymodel2D}
\end{figure*}

\section{testing \xs}
We now proceed to evaluate \xs~in a series of simulated velocity maps and real velocity fields.

\subsection{Toy model example} \label{sec:single_calse}
As an example of its use, we run \xs~on a simulated velocity map. This is the velocity field of a galaxy at 31.4 Mpc with a 32\arcsec~optical radius.
We model a velocity field with an oval distortion described by Eq.~\ref{bisymmetric}.
For the rotation curve we adopt the parameterization from \citet{Bertola1991}.
The non--circular motions were modeled using the Gamma probability density function; Gamma(2,3.5) for describing the $V_{2,t}$ component and Gamma(2,3) for $V_{2,r}$. The constant parameters were set to $\phi_{\mathrm{disk}}^{\prime} = 77^{\circ}$, $\phi_{\mathrm{bar}} = 35^{\circ}$, $i = 55^{\circ}$, $x_0 = 76.5 $, $y_0 = 75.2$ and $V_{sys} = 2142$~\kms. The field of view (FoV) is defined as $64\arcsec\times74\arcsec$ and the pixel scale was set to $0.5$\arcsec. Finally we convolved the image for decreasing its spatial resolution. We simulate a circular PSF with a 2D Gaussian function with a $1 \arcsec$ full width at half maximum (FWHM). 
We perturb the velocity profiles by adding Gaussian noise centered in zero and a standard deviation of $5$~\kms.
%No noise was added to the map since noise effects can not be distinguished from the blurring effects by the PSF.

We started \xs~by assigning random values for each of the constant parameters, except for $\phi_{\mathrm{bar}}$ which is initialized around its maximum (i.e., $\phi_{\mathrm{bar}} = 45^{\circ}$).
We set the initial and last ring exploration in $2.5\arcsec$ and $40\arcsec$ respectively; we also estimate the radial velocities each $2.5\arcsec$. A LS analysis was performed with 3 round iterations before starting the MCMC run. For comparison with Bayesian methods we adopt 1000 bootstraps during the LS analysis.
For MCMC sampling, we run a total of 4000 iterations with 60 different chains, which represents twice the number of free variables for this case. We discarded 50\% of the joint chain, for a total of 120k posterior samples on each parameter. The IAT for this run resulted in 127 which is superior to 50.

On the other hand, nested sampling only requires the prior information, for which we adopted the uniform priors from Table~\ref{Tab:priors}. No initial LS was performed for this case. We stopped the sampling procedure only when the remaining evidence to be integrated is $\leq 0.1$.

We test all the different kinematic models, i.e., circular, radial, bisymmetric model and we arbitrarily expand the harmonic series up to $M = 3$.
MCMC and NS derive the posterior distribution for each variable of the kinematic model; thus, we can take advantage of corner plots to represent their marginalized distributions and explore possible correlations between parameters. 
The median values and $2\sigma$ errors for each parameter are shown in Table~\ref{tab:res_simulation}, while in figure~\ref{fig:posteriors} we show the marginalized posteriors; for simplicity we only show the constant parameters for the bisymmetric model, although note this should be a $30\times30$ dimensions plot.

MCMC and nested sampling methods converge to the same solutions found by the LS method; this represents a great success for 
Bayesian methods to derive kinematic parameters from an input velocity map given the large dimension of the likelihood function.
We note that the input parameters are recovered in the bisymmetric model, not the case for the circular, radial and harmonic, as expected;
this is better appreciated in the corner plot from Figure~\ref{fig:posteriors}. MCMC and nested sampling recover the input parameters within the $1\sigma$ credible
interval, except for $y_0$ which lies  within $2\sigma$; among the constant parameters, the position angle of the oval distortion shows the larger uncertainty. 
From table~\ref{tab:res_simulation}, we notice that the uncertainties derived with Bayesian methods and bootstraps are of the same order.

The different velocities, $V_t(r)$, $V_{2,r}(r)$ and $V_{2,t}(r)$, are also recovered within the $2\sigma$ errors as observed from the rightmost panel from figure~\ref{fig:toymodel2D}. For consistence, in Appendix~\ref{sec:appendix_diskfit} we include the results using \texttt{DiskFit}; we notice that \xs~results are in total agreement with those obtained with \texttt{DiskFit}.

Table~\ref{tab:res_simulation} also shows the root mean square (RMS) for each kinematic model; models including non-circular rotation show a RMS value around 8.5 \kms. This leads to the question of which model is the preferred for describing a particular velocity field. In a statistical sense, when comparing 
different models one should choose the one with fewer parameters, since more variables in a model often reduce further the RMS, which does not necessarily provide the best physical interpretation of the data. 
Statistical tests such as the Bayesian information criterion (BIC) penalizes over
the variables from the model; BIC is defined in terms of the likelihood (or the chi-square) as, $\mathrm{BIC} = -2\ln \mathcal{L(\hat{\alpha})} + N_\mathrm{varys}\ln(N)$, where $\hat{\alpha}$ represents the parameters that maximize the likelihood function and $N$ is the number of data; thus, the model with the lowest BIC should be preferred.
Additionally, the evidence $\mathcal{Z}$ computed from NS, is a measure of the agreement of the data with the priors; in this way large (small) $\mathcal{Z}$ values are more (less) compatible with the priors.

However, when there is little information about the data, or only the data itself, it is difficult to select a model description of the data based on any information other than statistical tests. Regardless of the statistical method adopted, it is important to have a physical motivation for accepting or rejecting a model; otherwise, erroneous interpretations of the velocity field could arise. 

For the toy model example, Table~\ref{tab:res_simulation} shows that non-circular models have similar BIC values. Even when the harmonic decomposition model seems to perform a good fitting based on the residuals, the physical interpretation of the $m=2$ components are meaningless for this example.
Thus in a real scenario the radial and bisymmetric model should be compared. The Bayesian evidence for the radial and bisymmetric models results in $\ln \mathcal{Z} =-36136$ and $-36135$, respectively. In an statistically sense both solutions are equally probable; in other words, the data are insufficient for  making an informed judgment. This is not surprising given the simplicity of our velocity field model.

%Results using \diskfit~ are shown in Figure~\ref{fig:diskfit_res} from the Appendix section.

\subsection{Simulations}

\begin{figure}[t!]
\includegraphics[width = \textwidth]{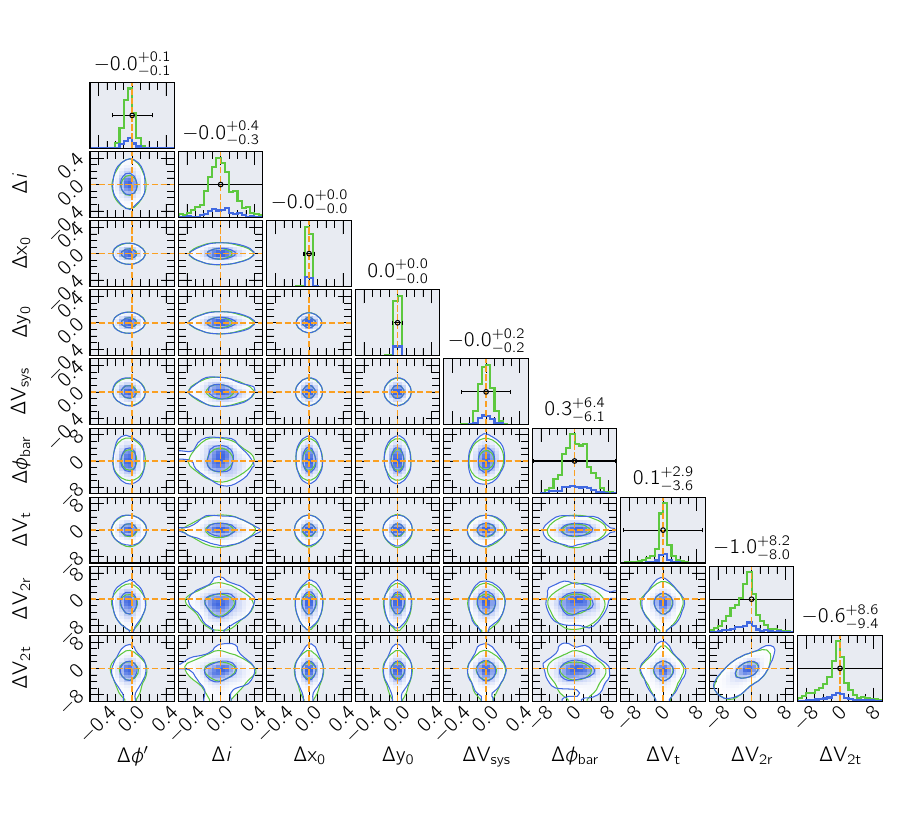}
\caption{\xs~results for the bisymmetric model on 1000 synthetic velocity maps with oval distortions. Values are reported respect their true value, namely $\Delta \alpha = \alpha_\mathrm{recovered}- \alpha_\mathrm{true}$, where $\alpha$ is any of the parameters considered in this example. Straight orange lines represent the true values (i.e., $\Delta \alpha=0$). Blue and lime colors show the results for NS and MCMC methods, respectively. The inner and outer density contours contain the 68\% and 95\% of the data, respectively. The upper histograms represent the 1D distributions of the parameters on the x-axis, while the size of the error bars represents the average value of the $2\sigma$ credible interval for each parameter. Note that  all parameters are recovered within the reported error bars. Values on-top the histograms represent the 50\%~percentile of $\Delta \alpha$, together with the $\pm 2 \sigma$ dispersion.}
\label{fig:1000sim}
\end{figure}

We carried out a set of 1000 simulations with different inclination angles ranging from  $30^\circ<i<70^\circ$, disk position angle $0^\circ<\phi_\mathrm{disk}^\prime<360^\circ$, and to avoid degeneracy, the bar position angle varies from $5^\circ<\phi_\mathrm{bar}<85^\circ$; velocity profiles and kinematic centre have the same values as in the toy example.
We also adopt the same sampling configurations as before. 
We notice that sometimes \xs~detects the minor-axis bar position angle instead of the major one; in such cases, \PAbar~is found $90^{\circ}$ away from the minor axis. This result is also a totally acceptable model since the difference resides only in the sign of the bisymmetric components, $V_{2,r}$ and $V_{2,t}$, which for these cases both have negative values. \xs~computes the projected major axis position angle of the bar via equation~\ref{eq:bar_projected}, while the projected minor axis position angle is computed by shifting \PAbar~by $90^{\circ}$.

In Figure~\ref{fig:1000sim} we show the results of the analysis in corner plots; the derived values are shown relative the true ones, namely $\Delta \alpha = \alpha_\mathrm{recovered}-\alpha_\mathrm{true}$; for the radial dependent velocities, we subtract the velocity profile of each component from the derived velocities.
These results show that the median values of $\Delta \alpha$ lie around zero for all parameters describing the bisymmetric model; furthermore, the scatter of the differences is contained within the average value of the $2\sigma$ credible interval for each parameter.
Results from this analysis demonstrate that MCMC and NS methods are cable to recover the true parameters of our simulated velocity maps,
even when each model is described by $+30$ free variables. 

The final accuracy of the recovered parameters would depend on the details of data themselves (resolution, \SN, spatial coverage etc.). Therefore, ad--hoc simulations are encouraged.

\subsection{NGC~7321}
\label{NGC7321}
We proceed to evaluate \xs~over the velocity field of a galaxy hosting a stellar bar. For this purpose we adopt the galaxy NGC~7321 observed as part of 
the CALIFA galaxy survey \citep[e.g.,][]{sanchez12a}. This object has been previously analyzed by \citet{Holmes2015} using \diskfit.
The \ha~velocity field of this object represents a good example of a galaxy with a strong kinematic distortion, most probably caused by the stellar bar. \citet{Holmes2015} found the bisymmetric model as the best model for reproducing the inner distortion observed in this object; they found best fit values and $1\sigma$ errors for the constant parameters given by \PAdisk = $12\pm1^{\circ}$, $i = 46\pm2^{\circ}$, Vsys = $7123\pm3$~\kms~and a bar position angle oriented at \PAbar $=47\pm6^{\circ}$. 
We implemented \xs~on the \ha~velocity map taken from the CALIFA dataproducts \citep[e.g.,][]{pipe3d}.
We adopted the same ring configurations as before, excluding pixels from the error map with values larger than 25~\kms; we proceed to explore the non-circular motions up to $r=18\arcsec$, and set the maximum radius for the circular velocities up to $40\arcsec$; this lead to a total of 36 free variables that will be estimated with Bayesian inference.
We adopt 3 rounds iterations for the LS method, and also compute the errors on the parameters with 1000 bootstraps.
For MCMC, we adopted 5000 steps and drop half of the total samples to let the joint chain stabilize. For NS we stop the sampling until the remaining evidence to be integrated is 0.1. 

\begin{figure}[t!]
\centering\includegraphics[width = 1\textwidth, trim = {0.5cm 0 0 0}, clip]{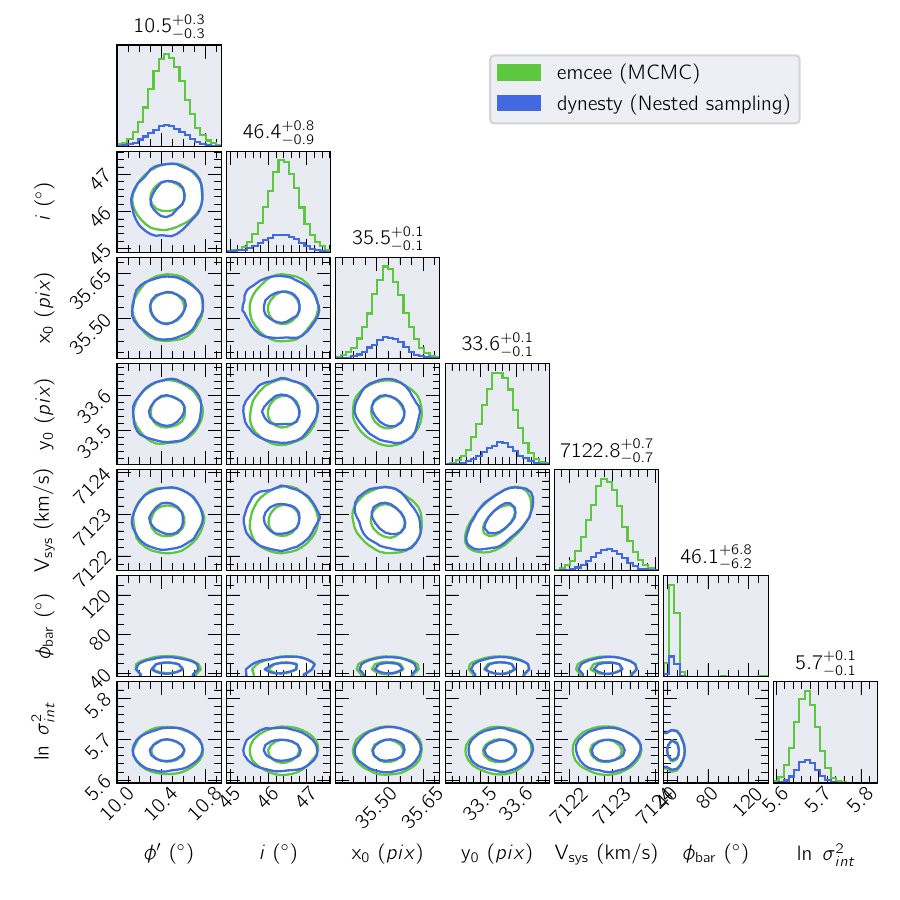}
\centering\includegraphics[width = 0.9\textwidth]{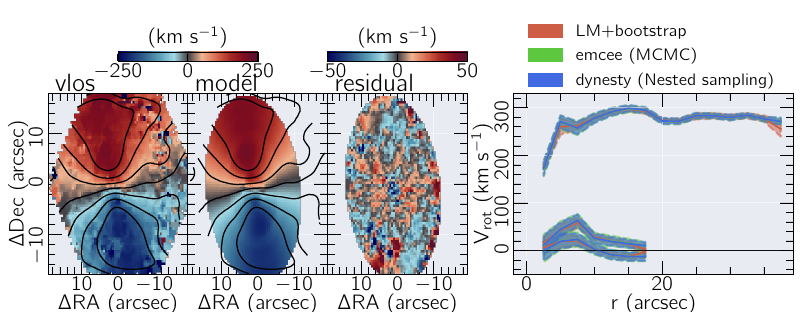}
\caption{\xs~results on NGC~7321 for the bisymmetric model. Top figure shows the marginalized distribution for the constant parameters with MCMC methods in green colors and NS in blue.  As in Figure~\ref{fig:posteriors}, the quoted values on top the histograms represent the median and 2$\sigma$ errors obtained from the posterior distributions. Bottom figure shows from left to right the \ha~velocity map; the two-dimensional model from NS; the residual map; and the radial profile of the different velocities. Shadow regions represent the $2\sigma$ credible intervals obtained from each method (bootstrap in red colors).}
\label{fig:corner_NGC7321}
\end{figure}

Figure~\ref{fig:corner_NGC7321} shows the marginalized distributions of the constant parameters. From the 1D histograms
we obtain \PAdisk = $11\pm0^{\circ}$,  $i = 46\pm1^{\circ}$, Vsys = $7123\pm1$~\kms~ and \PAbar $=46\pm6^{\circ}$; additionally we compute the intrinsic scatter of the data in $\sim16$~\kms.
Table~\ref{tab:NGC7321} shows a summary of these results. 
As can be read from this table, the constant parameters derived by \xs~are in concordance with those previously reported by \citet{Holmes2015}, although our uncertainties are smaller when comparing the errors at $2\sigma$, probably due to differences in methods.
The bottom figure shows the best model and residual map obtained from Nested sampling. The kinematic distortion observed in the central region  is well reproduced with the bisymmetric model. The bisymmetric motions, i.e., the bar-like flows, are oriented at $46\pm8^{\circ}$ on the sky plane. The rightmost panel shows the radial profile of the different velocity components derived from NS, MCMC and LS+bootstrap methods. Again, the uncertainties reported from NS and MCMC are of similar magnitude, and these are larger than those obtained with bootstrap methods.  

In Appendix~\ref{sec:other_implementations} we show the implementation of \xs~in other data with different instrumental configurations.

\begin{table}
\setlength{\tabnotewidth}{1\columnwidth}
\tablecols{7}
\caption{\xs~bisymmetric results for NGC~7321. }
\label{tab:NGC7321}
\begin{tabular}{lcccccc}
\toprule
{Method} & {\PAdisk} & {$i$} & {$x_0$}& {$y_0$} & {$V_\mathrm{sys}$} & {\PAbar}\\ 
 & ($^{\circ}$) & ($^{\circ}$) & (pixels) & (pixels) & (\kms) & ($^{\circ}$)\\
\midrule
LS+bootstraps & $11\pm1$ & $46\pm1$ & $35.5\pm0.0$ & $33.6\pm0.0$ & $7123\pm1$ & $46\pm4$\\
NS & $11\pm0$ & $46\pm1$ & $35.5\pm0.1$ & $33.6\pm0.1$ & $7123\pm1$ & $46\pm6$\\
MCMC & $11\pm0$ & $46\pm1$ & $35.5\pm0.1$ & $33.6\pm0.1$ & $7123\pm1$ & $46\pm7$\\
\texttt{DiskFit}\tabnotemark{*}& $12\pm1$ & $46\pm2$ & \nodata & \nodata & $7123\pm3$ & $47\pm6$\\
\bottomrule
\tabnotetext{*}{Results from \citet{Holmes2015}. Errors in \citet{Holmes2015} represent $1\sigma$, so a factor of 2 should be considered for comparison.}
\end{tabular}
\end{table}

\section{Discussion}
\label{sec:discussion}
Our simulations and toy example show that sampling methods are able to produce similar results as those obtained with frequentist methods based on the $\chi^2$ minimization. The widely used Levenberg-Marquardt algorithm is capable to obtain solutions to the kinematic models in a fast way, although errors from the covariance matrix result always small. 
On the other hand, our resampling implementation produces larger uncertainties compared with the covariance matrix. 
We notice that the magnitude of the errors increases with the number of bootstrap iterations; however, increasing the number of bootstrap samples increases the total execution time, since at each iteration a new LS analysis is performed. 

Bayesian methods, i.e., MCMC and NS, provide the largest uncertainties on the parameters among the two other techniques. The major disadvantage is the computational cost needed to sample the posterior distributions. For the toy model example presented, the execution times on an 8 core machine are $\sim1$~hour for LM+1k~bootstraps, $\sim1$~hour for MCMC  with 4k steps and $\sim2.5$~hours with NS.  

If Bayesian methods and LS+bootstrap provide similar solutions for the parameters, then in principle one could choose either of the two methods to quote the uncertainties. However, the most interesting cases are when Bayesian methods differ from the frequentist ones. 
\xs~has the advantage that both bootstrap and Bayesian methods can be executed in parallel. This can drastically reduce the execution times depending on the number of CPUs available during the running.

\section{Conclusions}
\label{sec:conclusions}
We have presented a tool for the kinematic study of circular and non--circular motions on galaxies with resolved velocity maps. This tool named \xs~(or \texttt{XS} for short), is an adaptation of the \texttt{DiskFit} algorithm, designed to perform Bayesian inference on parameters describing circular rotation, radial flows, bisymmetric motions and an arbitrary harmonic decomposition of the LoS velocities. \xs~implements robust Bayesian sampling methods to obtain the posterior distribution of the kinematic parameters. In this way, the ``best-fit'' values, and their uncertainties are obtained from the marginalized distributions, unlike frequentist methods where best values are obtained at a single point from the likelihood function.  
\xs~adopts Markov Chain Monte Carlo methods and dynamic nested sampling to sample the posterior distributions. In particular, \xs~makes use of the {\sc emcee} and {\sc dynesty} packages developed to perform Bayesian inference.

\xs~is a free access code written in Python language. The details about running the code, as well as the required inputs and the outputs are described in the Appendix~\ref{sec:appendix}.

\xs~is suitable to use on velocity maps not strongly affected by spatial resolution effects, i.e.,
when the PSF FWHM is smaller than the size of structural components of disk galaxies, such as stellar bars. In addition, disk inclination should range from $30^{\circ}<i<70^{\circ}$.
The fundamental assumption of the code is that galaxies are flat systems observed in projection in the sky with a constant inclination angle, constant disk position angle and fixed kinematic center. This makes \xs~suitable for studying the kinematics of galaxies within dynamical equilibrium, but not for highly perturbed disks.

%\xs~allows to explore the incidence of nonaxisymmetric flows in galaxies through radial, bisymmetric and a more general harmonic decomposition model. 

From applying \xs~over a set of simulated maps with oval distortions we showed that Bayesian methods are able to recover the input parameters despite the high dimension of the likelihood function. 
True parameters are recovered within $1\sigma$ credible interval, with the position angle of the oval distortion being the parameter with the larger scatter.
We tested \xs~over a well known galaxy with an oval distortion in the velocity field, NGC~7321, and found similar results to those obtained with \texttt{DiskFit}.

Regarding the uncertainty on the parameters, Bayesian methods provide the largest uncertainties compared with resampling methods like bootstrap. However, the computational cost for sampling the joint posterior distribution is in general more expensive than, for instance 1k bootstraps. Fortunately, a fraction of time can be saved when these methods are run in parallel.

We also tested \xs~ on velocity maps from different galaxy surveys.
Despite the instrumental differences in these data, \xs~is able to built kinematic models of circular and non-circular motions. 

Finally, \xs~is ideal for running on individual objects, or in galaxy samples since it is easy to systematize for use in large data sets. \xs~is a free access code available at the following link \url{https://github.com/CarlosCoba/XookSuut-code}. 

\section*{Acknowledgment}
We thank K. Spekkens, J. A. Sellwood, and anonymous peer reviewers for providing helpful suggestions to improve this manuscript.

C. L. C. thanks support from the IAA of Academia Sinica.
L. L. thank supports from the Academia Sinica under the Career Development Award CDA107-M03, the Ministry of Science \& Technology of Taiwan under the grant MOST 108-2628-M-001-001-MY3, and National Science and Technology Council under the grant NSTC 111-2112-M-001-044.

\newpage
\begin{appendices}
 
\section{Running XookSuut}
\label{sec:appendix}
\xs~is designed to run directly from the command line by passing a number of parameters that have for purpose guiding the user through a successful fit. %This makes \xs~ easy to automatize for the study of large galaxy samples. 

After a successful installation and typing {\tt XookSuut} on a terminal the code will display the entrance required for starting the analysis. The meaning of each parameter is described in Table~\ref{tab:entrance}, while the output files are described in Table~\ref{tab:dataproducts}.

\begin{longtable}{ p{2cm} p{1cm} p{9cm}}
\caption{\xs~input parameters.}\\
\hline
\textbf{Input} & \textbf{Type} & \textbf{Description} \\ 
\hline
name & str  &  Name of the object. \\
vel\_map.fits & fits  &  Fits file containing the 2D velocity map in \kms. \\
error\_map.fits& fits  &  Fits file containing the 2D error map in \kms. \\
SN & float  &  Pixels in the error map above this value are excluded. \\
pixel$\_$scale & float  & Pixel scale of the image (\arcsec /\ pixel). \\
PA & float  &  Kinematic position angle guess ($^{\circ}$). \\
INC & float  &  Disk inclination guess ($^{\circ}$). \\
X0 & float  &  X--coordinate of the  kinematic centre ($pix$). \\
Y0 & float  &  Y--coordinate of the  kinematic centre ($pix$). \\
VSYS & float  &  Initial guess for the systemic velocity in \kms. If no argument is passed, it will take the weighted mean value within a $5\arcsec$ aperture centered in (X0, Y0).\\
vary$\_$PA & bool  &  Whether \PAdisk~varies in the fit or not.\\ 
vary$\_$INC & bool  &  Whether $i$~varies in the fit or not.\\ 
vary$\_$X0 & bool  &  Whether $x_0$~varies in the fit or not.\\ 
vary$\_$Y0 & bool  &  Whether $y_0$~varies in the fit or not.\\ 
vary$\_$VSYS & bool  &  Whether V$_\mathrm{sys}$~varies in the fit or not.\\ 
ring$\_$space& float  &  Spacing between rings in arcsec. The user may want to use FWHM spatial resolution.\\
delta   & float &     The width of the ring is defined as 2delta. The user may want to use 0.5~ring$\_$space if independent rings are desired. \\
Rstart,Rfinal   & float & Starting and initial position of the rings on disk plane. (arcsec)\\
cover   & float & Fraction of pixels in a ring needed to compute the row stacked velocities. If 1 the ring area must be 100\% sampled.\\
kin\_model   & str  &  Choose between: ``circular'', ``radial'' flows, ``bisymmetric'' (oval distortion) or ``hrm\_M'', where M is the harmonic number.\\
fit\_method & str &  Minimization technique used in the Least-squares analysis. Options are ``Powell'' or ``LM'' (Levenberg--Marquardt). \\
N$\_$it & float  & Number of round iterations for the Least-squares analysis.\\
Rbar$\_$min,max   & float  &  Minimum and maximum radius for modeling the non--circular flows. If only one value is passed, it will be considered as the maximum radius to fit.\\
config$\_$file & file & Configuration file to access to high configuration settings including the Bayesian sampling methods, bootstrap errors, and other general model configurations. See the documentation for a detailed description of this file.\\
prefix & str &  Extra string passed to the object's name. This prevents overwriting the outputs in case of multiple analyses on the same object are performed. \\
\hline
\label{tab:entrance}
\end{longtable}

\begin{longtable}{ p{5cm} p{7cm}}
\caption{\xs~dataproducts}\\
\hline
\textbf{Output}  & \textbf{Description} \\
\hline
name.model.vlos\_model.fits.gz & Two dimensional representation of the adopted kinematic model (Eqs.~\ref{circular},\ref{radial},\ref{bisymmetric} or \ref{eq:harmonic}). \\
%\hline
name.model.chisq.fits.gz & Fits file containing the chi-square map defined as  (obs-model)$^2$/error$^2$ \\
%\hline
name.model.chain.fits.gz & Fits file containing the marginalized samples (i.e., the joint chain), explored in the Bayesian analysis.\\
%\hline
name.model.2D\_Vmodel.fits.gz & Two dimensional representation of each velocity component from the model. \\
%\hline
name.model.marginal\_dist.fits.gz & Fits file containing the 50 percentile distribution for each parameter, together with the $\pm1\sigma$ and $\pm2\sigma$ credible intervals. \\
%\hline
name.model.residual.fits.gz & Map containing the residuals of the model, i.e., obs - model. \\
%\hline
name.model.2D\_R.fits.gz & Deprojected distance map in arcsec, obtained from the best fit disk geometry. \\
%\hline
name.model.1D\_model.fits.gz & Values for the best fit parameters together with the $2\sigma$ errors.\\
name.model.2D\_theta.fits.gz & Two dimensional representation of the azimuthal angle $\theta$.\\
\hline
\label{tab:dataproducts}
\end{longtable}

\section{Cauchy distribution}
\label{sec:cauchy}
\begin{figure}[t!]
\centering\includegraphics[width = 0.9\textwidth]{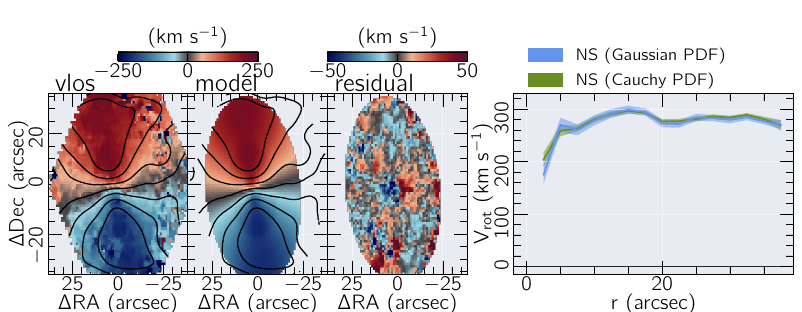}
\caption{Results for Gaussian and Cauchy likelihood functions for the circular model on NGC\,7321. In this example we used Nested sampling for the Bayesian analysis. We found the width of the Cauchy distribution in $\gamma = 7.6\pm0.3$~\kms. }
\label{fig:cauchy}
\end{figure}

Although Gaussian distribution is the most assumed for the likelihood function, there is no
restriction to use other distributions. In fact, multiple algorithms adopt arbitrary parameterization of the
residual function \citep[e.g.,][]{3Dbarolo}.
In addition to Gaussian distribution, \xs~also includes the Cauchy distribution in the likelihood function. It assumes a unique form of the errors parameterized with $\gamma$. The Cauchy log-posterior distribution for our models adopts the following form:

\begin{multline}
\label{eq:cauchy}
%log_L = -(N + 1) * np.log(beta) - np.sum(np.log(1.0 + ( dy / beta )**2) )
  \ln p(\vec{\alpha} | \mathcal{D},\mathcal{V}_\mathrm{model}) =  -N\ln \pi \gamma - \sum_{n=1}^N \ln \Big( 1 + \frac{ \big( \mathcal{D}_n - \sum_{k=1}^K W_{k,n} \overrightarrow V_k \big)^2 }{\gamma^2} \Big)  \\
           + \ln p(\alpha) - \ln \mathcal{Z}
\end{multline}
An example using the Cauchy distribution is shown in Figure~\ref{fig:cauchy}.
As noted, there can be differences in the results depending on the election of the likelihood function.
There is no a general rule on when to use the Cauchy distribution; often, it is used when the data contain many outliers.

\section{\diskfit~results}
\label{sec:appendix_diskfit}
\begin{figure}[t]
\centering
\includegraphics[scale=2]{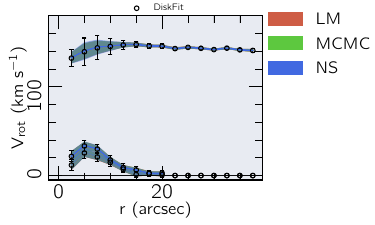}
\caption{Results of the fitting analysis using \diskfit~for the simulated velocity map described in Sec~\ref{sec:single_calse}. Black empty circles and error bars show velocities and uncertainties derived by \diskfit; 1000 bootstraps were adopted in \diskfit~for this purpose. Colored lines represent results from \xs~using LM+bootstrap in red, MCMC in green and NS in blue. Results for the constant parameters using \diskfit~are the following, \PAdisk~$=77.02\pm0.13^{\circ}$, $i = 55.06\pm0.48^{\circ}$, $(x_0,~y_0) = (77.51\pm0.09,~76.26\pm0.09)$ pixels, $V_\mathrm{sys} = 2141.98\pm0.20$~\kms~and \PAbar~ = $39.47\pm15.60^{\circ}$, $\chi^2=72.5$.  In all cases error bars represent $2\sigma$ errors.}
\label{fig:diskfit_res}
\end{figure}

Figure~\ref{fig:diskfit_res} shows the results of the bisymmetric model using \diskfit~applied on the simulated velocity map described in Section~\ref{sec:single_calse}; 1000 bootstraps were adopted in \texttt{DiskFit} to quote the uncertainties on the parameters.
The median values estimated with Bayesian methods and LM+bootstraps are in concordance with those obtained with \diskfit. In addition, the uncertainties on the velocities reported by \diskfit, are comparable or  lower than to those obtained with Bayesian methods. This figure shows that \xs~produces similar results as \diskfit.

\section{Implementation on data with different configurations}
\label{sec:other_implementations}
We apply \xs~to a sample of galaxies observed with different instrumental configurations and different redshifts. We obtain \ha~velocity maps from different integral field spectroscopy (IFS) galaxy surveys, namely  MaNGA \citep[e.g.,][]{MaNGA}, AMUSING++ \citep[e.g.,][]{AMUSING++}, SAMI \citep[e.g.,][]{SAMI}; these objects correspond to manga-9894-6104, IC\,1320 and SAMI511867, respectively. These objects were chosen for showing rich emission in \ha. The velocity maps were obtained from the public dataproducts. 

For each galaxy we run circular, radial, bisymmetric and harmonic decomposition model up to $M=3$. However we only report the model with the lowest BIC value. The initial position angle and inclination angles were adopted from those reported in Hyperleda or by own previous analysis \citep[e.g.,][]{walcher14,AMUSING++}. 
The coordinates of the kinematic center were set by eye from the velocity maps.
When available we use the error maps to exclude pixels with large uncertainties (namely  $>$ 25~\kms). The width of the rings was set to the size of the PSF for each dataset (ranging from      $1\arcsec-2.5\arcsec$). 
For these objects we only adopt NS methods. For speeding up the analysis we adopt truncated Gaussian priors, for which we perform a LS analysis to set the mean values of the Gaussian priors. 

\begin{table}\centering
\small
\setlength{\tabnotewidth}{0.5\columnwidth}
\tablecols{8}
\caption{\xs~applied on different data with different instrumental configurations. }
\label{tab:results_modeling_ifs}
\begin{tabular}{lccccccc}
\toprule
{Object} & {Survey} & {Model} & {\PAdisk} & {$i$} & {$x_0$}& {$y_0$} & {$V_\mathrm{sys}$}\\ 
 &   &   & ($^{\circ}$) & ($^{\circ}$) & (pixels) & (pixels) & (\kms) \\
 \midrule
9894-6104 & MaNGA & Eq~\ref{circular} & 296 $\pm$ 0 & 35 $\pm$ 1 & 27.2 $\pm$ 0.1 & 26.7 $\pm$ 0.1 & 10696 $\pm$ 1\\
511867 & SAMI  & Eq~\ref{radial} & 206.0 $\pm$ 2 & 46 $\pm$ 1 & 24.9 $\pm$ 0.1 & 24.5 $\pm$ 0.2 & 16493 $\pm$ 1\\
IC\,1320 & AMUSING++  & Eq~\ref{eq:harmonic} & 85 $\pm$ 0 & 57 $\pm$ 3 & 165.4 $\pm$ 0.1 & 168.7 $\pm$ 0.0 &  4950 $\pm$ 0\\
\bottomrule
\end{tabular}
\end{table}
%%%%

\begin{figure*}[t!]
\centering
\includegraphics[width = \textwidth]{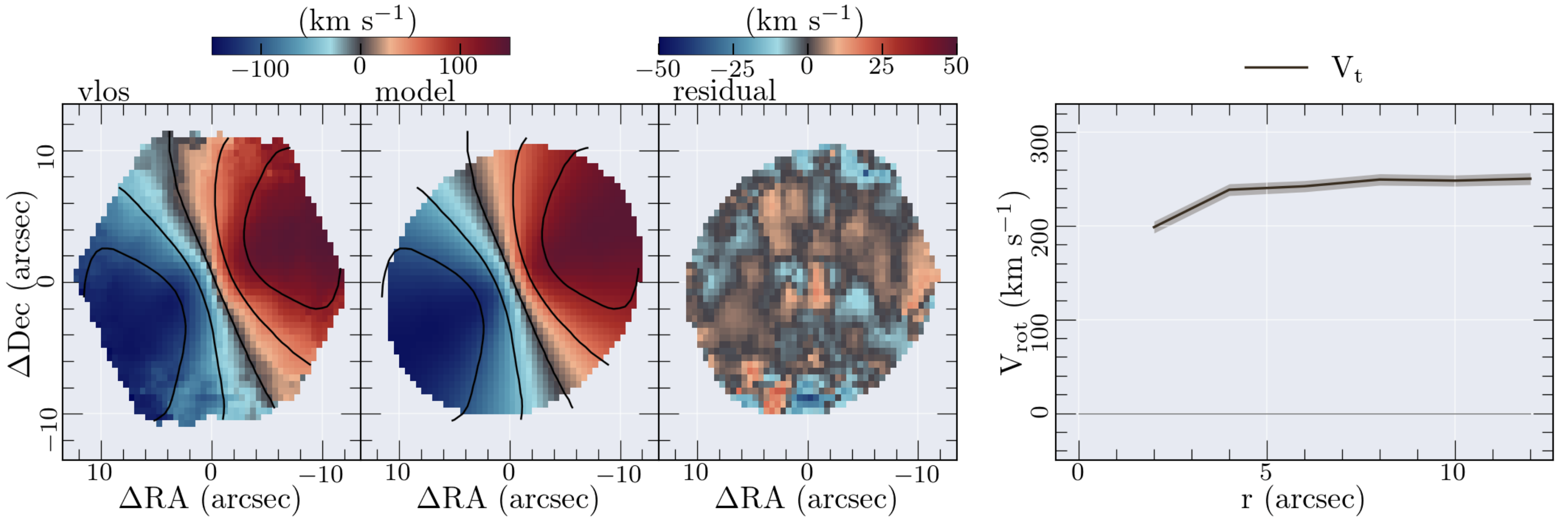}
\includegraphics[width = \textwidth]{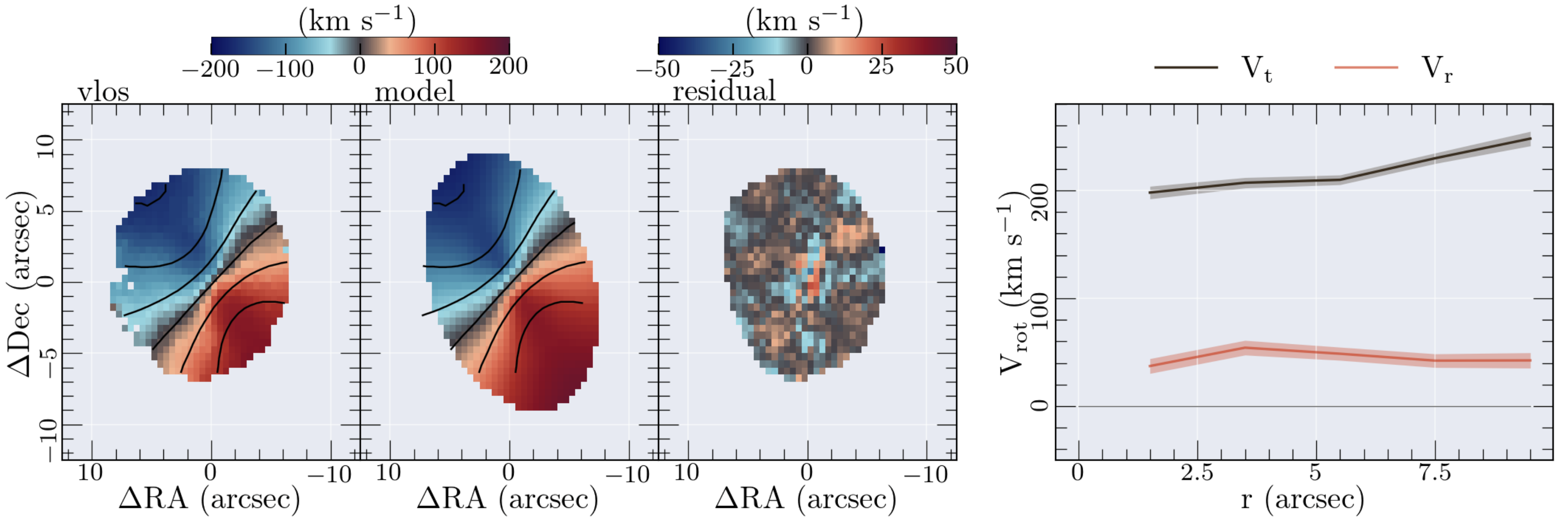}
\includegraphics[width = \textwidth]{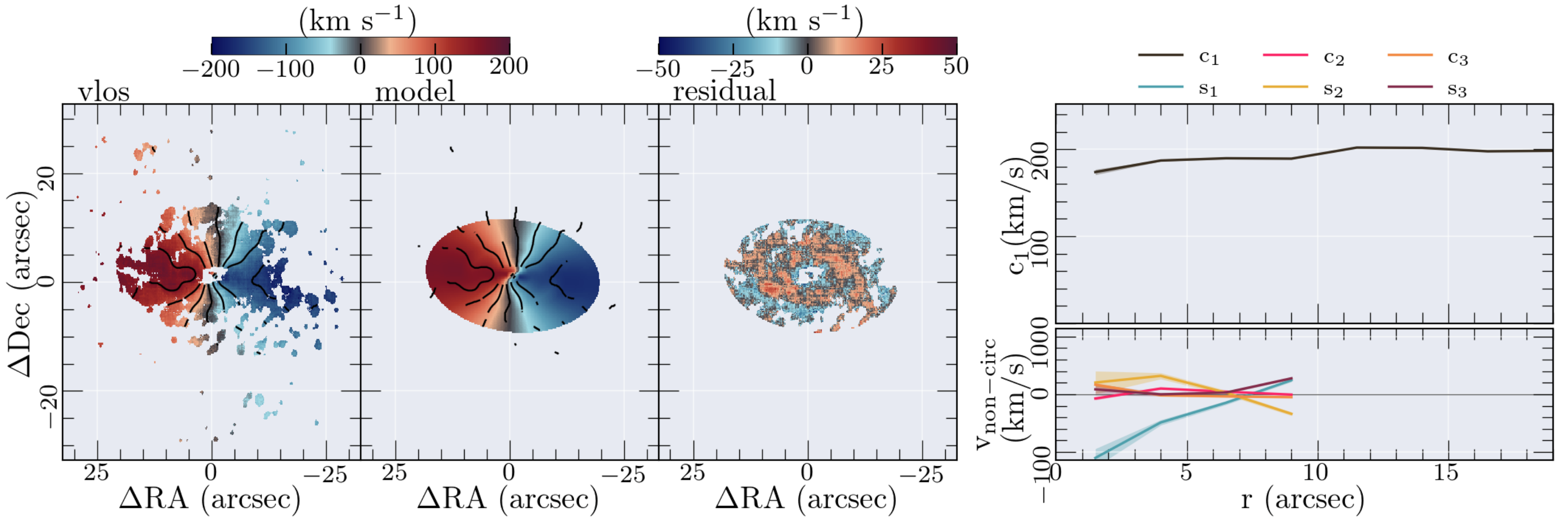}
\caption{Implementation of \xs~on different velocity maps. Each set of panels, from top to bottom, correspond to a different galaxy taken from different IFS galaxy surveys (i.e.,  MaNGA, SAMI and AMUSING++ from top to bottom). In each row, from left to right: (i)  the \ha~velocity field; (ii) best two--dimensional model from NS; (iii) residual map of the fitting; (iv) radial variation of the different velocities in the considered model. Shadow regions in this plot represent the $1\sigma$ credible interval obtained from NS. Note that each map has different instrumental configurations and FoVs. Iso-velocity contours spaced by $\pm50$~\kms~are overlayed in each velocity map.}
\label{ifs_kin_mod}
\end{figure*}
The best fit models are shown in Figure~\ref{ifs_kin_mod}, while results of the constant parameters are shown in Table~\ref{tab:results_modeling_ifs} for each object.
Figure~\ref{ifs_kin_mod} shows the observed velocity, the best model from NS methods, the residual maps and the radial profiles of the different kinematic components for each considered model. Each row in this figure represents the outputs for a different galaxy.

The manga--9894--6104 galaxy is well described by the circular model. It shows a symmetric velocity field with orthogonal major and minor axes. The circular model describes well the observed velocities and produce small residuals of the order of  $\pm10$~\kms. The rotation curve is flat within the FoV, with $V_\mathrm{max} \sim 248$~\kms.

The velocity field of SAMI511867 shows a slight twist along the minor axis, which is well reproduced by the radial flow model.
Significant contribution of radial motions of the order of $40$~\kms~ are observed across the SAMI FoV. However, because of its small FoV , large PSF $\sim2^{\prime\prime}$ and the physical spatial resolution (FWHM $\sim$ 2~kpc), parameters derived in Table~\ref{tab:results_modeling_ifs} could be affected by these effects. 

Finally, IC\,1320  is part of the AMUSING$++$ galaxy compilation. This object was observed with the modern instrument MUSE \citep[e.g.,][]{MUSE}. The IFU of this instrument has the smaller spaxel size ($0.2\arcsec$) and the best spatial resolution (seeing limited) from the data analyzed here; as a consequence, IC\,1320  shows a velocities field rich in details.  Among the different kinematic models, the harmonic model showed the lowest BIC value  The $c_1$ component, which is a proxy of the circular rotation, is mostly flat across its optical extension with $v_\mathrm{max}\sim 200$~\kms. Non--circular terms are dominant within the inner $10\arcsec$. The $c_3$ and $s_3$ coefficients may indicate the presence of stream flows associated with spiral arms or a stellar bar.
 
\end{appendices}
\bibliographystyle{rmaa}
%\bibliography{bibliography} % if your bibtex file is called example.bib

\end{document}